\documentclass[aps,prl,reprint,longbibliography]{revtex4-2}
\pdfoutput=1 

\usepackage[scaled]{helvet}
\usepackage{graphicx}
\usepackage{caption}
\usepackage{amsmath,amsthm,amsfonts,amssymb}
\usepackage{graphicx}
\usepackage{marginnote}
\usepackage{float}
\usepackage{color}
\usepackage{array}
\usepackage{multirow}
\usepackage[normalem]{ulem}
\usepackage{upgreek}
\usepackage{wasysym}

\usepackage{url}
\usepackage{rotating}
\usepackage{gensymb}
\graphicspath{{./figures/}}
\usepackage{hyperref}
\usepackage{ulem}
\usepackage[english]{babel}
\usepackage[T1]{fontenc}
\usepackage[utf8]{inputenc}
\usepackage[table]{xcolor}
\usepackage{xr}

\renewcommand{\tensor}[1]{{\mathsf{\bf#1}}}
\renewcommand{\vec}[1]{{\boldsymbol{#1}}}

\def\X{\bar{\tensor{X}}}

\def\b{\boldsymbol{\beta}}

\def\D{\tensor{D}}
\def\g{\boldsymbol{\gamma}}
\def\mG{m_{\scriptscriptstyle\textrm{GA}}}

\def\Tgval{T_{\! \textrm{g}}}
\def\Tga{T_{\! \textrm{g}}^{\,a}}
\def\Tgahat{\hat{T}_{\! \textrm{g}}^{\,a}}
\def\Tgbhat{\hat{T}_{\! \textrm{g}}^{\,b}}
\def\Tgvec{\boldsymbol{T}_{\!\! \textbf{\textrm{g}}}}

\newcommand{\taualpha}{\tau_{\alpha}}

\mathchardef\mhyphen="2D

\newif\ifmarkup
\markuptrue 
\ifmarkup
\newcommand{\delete}[1]{{\color{red}\sout{#1}}}

\newcommand{\pdonote}[1]{{\color{blue}{[#1]}}}

\newcommand{\scnote}[1]{{\color{cyan}{[#1]}}}

\newcommand{\jm}[1]{{\color{magenta}{[#1]}}}
\else
\newcommand{\delete}[1]{}

\newcommand{\pdonote}[1]{}

\newcommand{\scnote}[1]{}
\newcommand{\jm}[1]{}\fi

\begin{document}
\renewcommand{\thefootnote}{\fnsymbol{footnote}}
\renewcommand{\:}{\!\!:\!\!}

\title{A fast transferable method for predicting the glass transition temperature of polymers from chemical structure}

\author{Sebastian Brierley-Croft$^1$} 
	\author{Peter D. Olmsted$^3$}
 	\author{Peter J.~Hine$^1$} 
    \author{Richard J.~Mandle$^{1,2}$} 
    \author{Adam Chaplin$^4$}
    \author{John Grasmeder$^4$} 
	\author{Johan Mattsson$^1$}
    \email{k.j.l.mattsson@leeds.ac.uk}
	\affiliation{$^1$School of Physics and Astronomy, University of Leeds, Leeds LS2\,9JT, United Kingdom}
        \affiliation{$^2$School of Chemistry, University of Leeds, Leeds LS2\,9JT, United Kingdom}
	\affiliation{$^3$Department of Physics and Institute for Soft Matter Synthesis and Metrology, Georgetown University, Washington DC, 20057}
	\affiliation{$^4$Victrex PLC, Hillhouse International, Thornton Cleveleys, Lancashire FY5 4 QD, United Kingdom}
	\date{\today}

\begin{abstract}
We present a new method that successfully predicts the glass transition temperature $\Tgval$ of polymers based on their monomer structure. The model combines ideas from Group Additive Properties (GAP) and Quantitative Structure Property Relationship (QSPR) methods, where GAP (or Group Contributions) assumes that sub-monomer motifs contribute additively to $\Tgval$, and QSPR links $\Tgval$ to the physico-chemical properties of the structure through a set of molecular descriptors. This method yields fast and accurate predictions of $\Tgval$ for polymers based on chemical motifs \emph{outside} the data sample, which resolves the main limitation of the GAP approach. Using a genetic algorithm, we show that only two molecular descriptors are necessary to predict $\Tgval$ for PAEK polymers. Our QSPR-GAP method is readily transferred to other physical properties, to measures of activity (QSAR), or to different classes of polymers such as conjugated or bio-polymers.

\end{abstract}

\maketitle

\section{Introduction}

Polymers are remarkably versatile materials, and the combined control of monomer chemistry and chain length allows for superior tuneability of physical properties. As a polymer melt is cooled, the time-scale $\taualpha$ characterising its structural ($\alpha$) relaxation increases dramatically, and in the absence of crystallisation the structure freezes into an amorphous solid, a glass, at the glass transition temperature $\Tgval$ \cite{angell2000relaxation}. Since molecular motions are controlled by $\Tgval$, this is a key parameter for understanding and predicting material behaviour, and it is thus essential to develop methods for accurately predicting $\Tgval$ directly from the chemical structure.

For long-chain polymers, $\Tgval$ is molecular weight ($M$) independent \cite{fox1950second,cowie1975some,novikov2003universality,Baker2022PRX}, but strongly affected both by intramolecular dihedral barriers \cite{bernabei2008dynamic,colmenero2015polymers} (chain flexibility) and intermolecular packing effects, both of which are chemistry-specific \cite{Baker2022PRX}. Importantly, it has been shown that the $\alpha$ relaxation, which defines $\Tgval$, is linked to relaxations on a relatively `local' sub-monomer length-scale \cite{Boyer1963RubbChemTech,Bershtein1985JPolSciPolymLettEd,Boyer1976polymer,Boyd1985Polymer,Boyd1985Polymer2,Ngai1985JPolSciPolPhysEd,Roland2005RepProgPhys}, which in turn suggests that models that predict $\Tgval$ from monomer structure should be achievable. In this paper we present such a model, and apply it to the poly(aryl ether ketone) (PAEK) family of polymers.

Predictive models that relate structure-based properties and $\Tgval$, and are suitable for small data sets with low chemical variability, have been proposed for polymers \cite{Schneider1992Thedata, matsuoka1997entropy, Schneider2005Polymertemperature, Schut2007GlassPrinciple, Xie2020GlassPolymers, Baker2022PRX, Alesadi2022MachineStructure, Pilania2019Machine-Learning-BasedCopolymers}. For instance, an approximate correlation has been found between $\Tgval$ and monomer-scale properties such as the molecular weight per conformational (or flexible) degree of freedom of the chain ($M_{\phi}$) \cite{matsuoka1997entropy,Schut2007GlassPrinciple,Baker2022PRX,Schneider1992Thedata, Schneider2005Polymertemperature}, where $M_{\phi}$ captures both chain flexibility and chain bulkiness (reflecting molecular packing). As one example, \citet{Schut2007GlassPrinciple} correlated $\Tgval$ with the mass per flexible bond for a data set divided into three polymer classes by introducing flexible groups into both the main chain and the side chains; an out-of-sample mean absolute error (MAE) for $\Tgval$ of $\lesssim$ $6\,\textrm{K}$ (per polymer class) was obtained.
In another example, \citet{Xie2020GlassPolymers} assigned an ad-hoc mobility factor to each atom based on the chemical group it belongs to (\textit{e.g.}, alkyl, phenyl or thiophene). The monomer's mobility was then averaged over the atomic contributions, followed by a regression of $\Tgval$ on the monomer mobility. For a family of 32 conjugated polymers, a RMSE $\simeq 13\,\textrm{K}$ was attained for in-sample $\Tgval$ predictions. These methods are easily applicable and intuitive; \textit{e.g.} by linking a relevant physical property, such as a molecular weight or volume, to each `flexible bond', where ad-hoc rules are often introduced to quantify the influence of different bonds. However, the approaches are typically tailored to specific data sets and are not generalisable to a wider set of polymer structures \cite{Alesadi2022MachineStructure}.

Conversely, a more generalisable approach is the so-called group contribution, or group additive properties (GAP) method \cite{Boyer1963ThePolymers,Weyland1970PredictionPolymers,vanKrevelen1990PropertiesPolymers}. It assumes that a polymer property can be expressed by a composition-weighted average over contributions from sub-monomer motifs (fragments). The fragment contributions can be determined directly from data by a linear regression.
\citet{vanKrevelen1990PropertiesPolymers} applied GAP to predict various polymer properties, such as transition temperatures, solubility, mechanical, optical and electrical properties; while \citet{Weyland1970PredictionPolymers} quoted in-sample MAE $\simeq 10\,\textrm{K}$ for predictions of $\Tgval$. Despite their broad applicability, a fundamental flaw of GAP models is that they cannot be used to make predictions for polymers containing fragments outside of the data sample \cite{Katritzky1998B,Hopfinger1988MolecularTemperatures,Pilania2019Machine-Learning-BasedCopolymers}.

A method that addresses some shortcomings of GAP models is the so-called quantitative structure-property relationship (QSPR) approach. QSPR-based methods use molecular descriptors \cite{Todeschini2000HandbookDescriptors, Todeschini2009MolecularChemoinformatics}, which quantify electronic, topological or geometric properties that are calculated from atomistic representations of molecules. For polymers, QSPR methods are normally applied either to the monomer \cite{Pilania2019Machine-Learning-BasedCopolymers, Bejagam2022MachineBiopolymers, Katritzky1998B, Le2012QuantitativeProperties} or to oligomers consisting of a few monomers \cite{SanchezLengeling2019AParameters, ParandekarModelingRelationships2015, DuchowiczQSPRPolymers2015}, and statistical or machine learning (ML) techniques are used to determine the relationship between the descriptors and the investigated property (such as $\Tgval$) \cite{Gasteiger2003HandbookChemoinformatics, Katritzky2010QuantitativePrediction, Le2012QuantitativeProperties}. For QSPR methods applied to $\Tgval$ predictions, RMSEs typically vary from $\simeq 4-35\,\textrm{K}$ 
\cite{Pilania2019Machine-Learning-BasedCopolymers, Kim2018PolymerPredictions, Jha2019Impacttemperatures, Katritzky1998B}, 
depending on the chemical variation within the data set. Models on larger data sets \cite{Huan2016ADesign, Otsuka2011PoLyInfo:Design}, with higher chemical variation, typically yield prediction errors exceeding 25~K \cite{Kim2018PolymerPredictions, Kuenneth2023PolyBERT:Informatics}.
A significant drawback of QSPR models is that accurate descriptor calculations can be computationally costly, especially for large monomers or oligomers.

GAP and QSPR methods have usually been applied separately \cite{Pilania2019Machine-Learning-BasedCopolymers, Bejagam2022MachineBiopolymers, Xie2020GlassPolymers, Alesadi2022MachineStructure}. However, \citet{Hopfinger1988MolecularTemperatures} proposed a linear regression-based model for predicting $\Tgval$ based on a GAP-like averaging scheme, combined with associating physical properties (conformational entropy and mass) with individual bonds. Inspired by this approach, we suggest extending QSPR methods to a smaller structural scale than the monomer unit, assuming interactions between these sub-monomer motifs negligibly contribute to the property of interest.

Here, we resolve the shortcomings of both the GAP and standard QSPR models, by developing a hybrid QSPR-GAP method: a molecule is divided into sub-monomer fragments for which molecular descriptors are calculated, and various linear regression methods are used to link $\Tgval$ to the fragment structure. Our approach significantly accelerates the descriptor calculations and addresses the weakness of GAP methods, also providing accurate predictions for polymers containing fragments outside the data sample. 

\begin{figure*}[htbp]
\includegraphics[width=\textwidth]{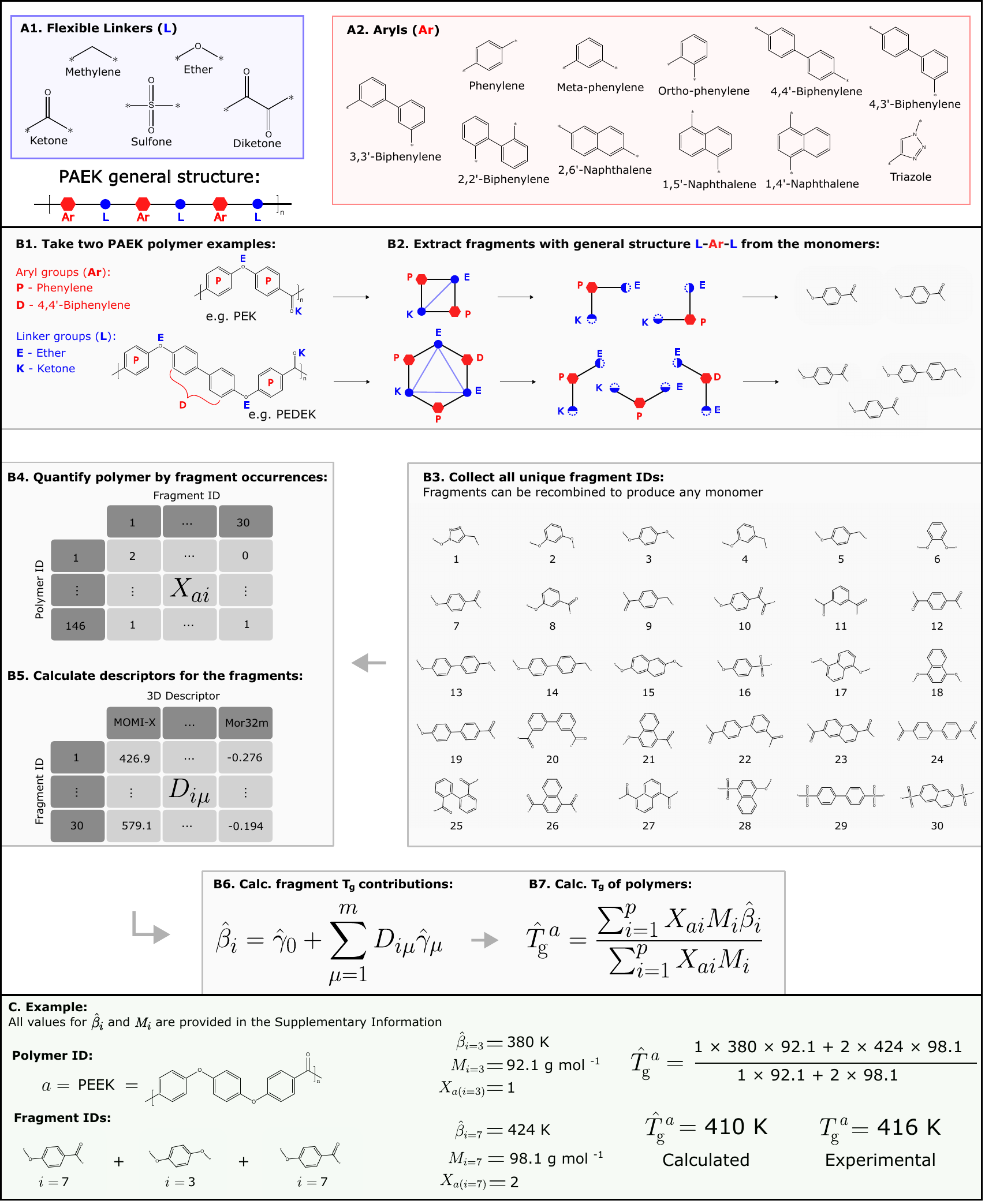}
\caption{\textbf{How to predict $\Tgval$ from the monomer structure of PAEK polymers.} \textbf{A}~The chemical building blocks (flexible linker and aryl moieties) of the PAEK polymer data set. \textbf{B}~Step-by-step illustration of the process by which $\Tgval$ is predicted from the monomer structure. \textbf{C}~An example calculation performed on poly(ether ether ketone) (PEEK) where $\hat{\beta}_i$ values determined from the QSPR-GAP Lasso model are used. All values for $\hat{\beta}_i$ and $M_i$ are listed in Table~SV; SI.}
\label{fig:Flow_diagram}
\end{figure*}

We apply our new QSPR-GAP method to a data set of 146 linear homo- and copolymers of poly(aryl ether ketone) (PAEK) -- an important class of linear polymers characterised by alternating stiff (aryls such as phenyls or biphenyls) and flexible linker (such as ethers or ketones) moieties, as shown in Fig.~\ref{fig:Flow_diagram}-A. The properties of PAEK polymers are highly tuneable by varying these moieties, making them suitable for a wide range of applications including smart-phone speakers, electrical insulation, automotive gears, medical implants and aircraft components \cite{Kemmish2010UpdatePolyaryletherketones}. To design PAEK polymers with optimised properties for specific applications, reliable structure-property relationships are essential.

We use our QSPR-GAP method to predict $\Tgval$ from the monomer structure, with an RMSE\,$\simeq 5-12\,\textrm{K}$ (out-of-sample). In cases where the GAP model is known to fail (\textit{i.e.}, predicting polymers containing fragments outside of the training set), the model makes accurate  predictions.
Moreover, by identifying the molecular descriptors most important for predicting $\Tgval$, we reach new insights into how local
molecular structure relates to the glass transition temperature in polymers. Our findings offer a pathway to predict the properties of highly complex polymer structures using small data sets, thus circumventing the need for more elaborate ML methods, which typically require larger data sets. Our method is readily generalisable to both a wider range of polymer properties (such as mechanical, optical, or electrical properties), and different classes of polymers.

\section{Results and Discussion}

\subsection{Characterisation of PAEK polymers}
Our QSPR-GAP model is applied to a data set of 77 PAEK homopolymers and 69 copolymers, sourced from both the literature and from experimental measurements conducted by Victrex R\&D. We ignore any minor effects of chain length on $\Tgval$ \cite{Baker2022PRX} and assume that all measured $\Tgval\equiv \Tgval^{\infty}$ (the long chain limit). 

The monomer of a PAEK polymer (see examples in Fig.~\ref{fig:Flow_diagram}-B1) is a sequence of alternating rigid aryl $Ar$ (Fig.~\ref{fig:Flow_diagram}-A2) and flexible linker $L$ (Fig.~\ref{fig:Flow_diagram}-A1) moieties, where the alternating arrangement
\begin{equation}
    \ldots L_1\-Ar_1\-L_2\-Ar_2\ldots L_N\-Ar_N \ldots
\end{equation}
is simple, yet different choices of $Ar$ and $L$ moieties lead to diverse material behaviour, as illustrated by the $\Tgval$ range of 375-550 K for the present polymer data set (Fig.~S1; SI). 

We divide the monomer structure into unique sub-monomer `fragments' that constitute all PAEK monomers in the data set. Many fragment choices are possible, including: $L\-Ar$, $L\-Ar\-L$, or $Ar\-L\-Ar$, or even longer sections. However, we mainly focus on $L\-Ar\-L$, since the calculation of descriptors (see details below) requires the addition of hydrogens to the two ends of the fragments, and $L\-Ar\-L$ is the only candidate that retains the uniqueness of the fragments once end-capped with hydrogens (Fig.~S20; SI). The data set of 146 polymers comprises 30 unique $L\-Ar\-L$ fragments  (Fig.~\ref{fig:Flow_diagram}-B3) and as two examples, Fig.~\ref{fig:Flow_diagram}-B1 illustrates how the monomers of poly(ether ketone) (PEK) and poly(ether biphenyl ether ketone) (PEDEK) are divided into $L\-Ar\-L$ fragments. 

Each homopolymer is parameterised by its \textit{count matrix} $\tensor{X}$, where $X_{ai}$ is the (integer) number of occurrences of fragment $i$ in homopolymer $a$'s monomer (see illustration in Fig.~\ref{fig:Flow_diagram}-B4). Correspondingly, for copolymer $a$ we define $X_{ai} = \sum_{\xi=1}^l w_\xi X_{\xi ai}$, where $w_\xi$ is the molecular weight fraction of comonomer $\xi$, and $X_{\xi ai}$ is the count of fragment $i$ in copolymer $a$'s comonomer $\xi$.

The molecular descriptors for each monomer fragment are determined as follows: (i) Add hydrogen atoms to the ends of each fragment $i$ (Fig.~\ref{fig:Flow_diagram}-B3); (ii) generate an energy-minimised 3D representation of each fragment using the Merck Molecular Force Field (MMFF) \cite{Halgren1996MerckMMFF94} with \texttt{RDKit} \cite{RDkit}; (iii)~calculate the values of molecular descriptors using \texttt{Mordred} \cite{Moriwaki2018Mordred:Calculator}. We calculate $m=213$ descriptors for each of the $p=30$ fragments. The descriptors consist of six types: 41 charged partial surface area (CPSA) descriptors, 4 geometrical indices, 4 gravitational indices \cite{Todeschini2000HandbookDescriptors,Todeschini2009MolecularChemoinformatics}, 160 3D-MoRSE descriptors \cite{Schuur1996TheActivity, Schuur1997InfraredRepresentation, Devinyak20143D-MoRSEExplained}, 3 moment of inertia descriptors, and 1 plane of best fit (PBF) descriptor \cite{Firth2012PlaneMolecules}. These $\mu=1, 2, \ldots, m$ descriptors encoding the $i=1,\ldots,p$ fragments constitute a \textit{descriptor matrix} $\tensor{D}$, where $D_{i\mu}$, provides the value of descriptor $\mu$ for fragment $i$ (see Fig.~\ref{fig:Flow_diagram}-B5).

The process of calculating descriptors for only $p=30$ sub-monomer fragments significantly reduces computation compared with descriptor calculations for the $n=146$ monomer units (the standard polymer QSPR approach) \cite{Katritzky2010QuantitativePrediction, Le2012QuantitativeProperties}.

\begin{figure*}[htbp]
\begin{center}
\includegraphics[width=1\textwidth]{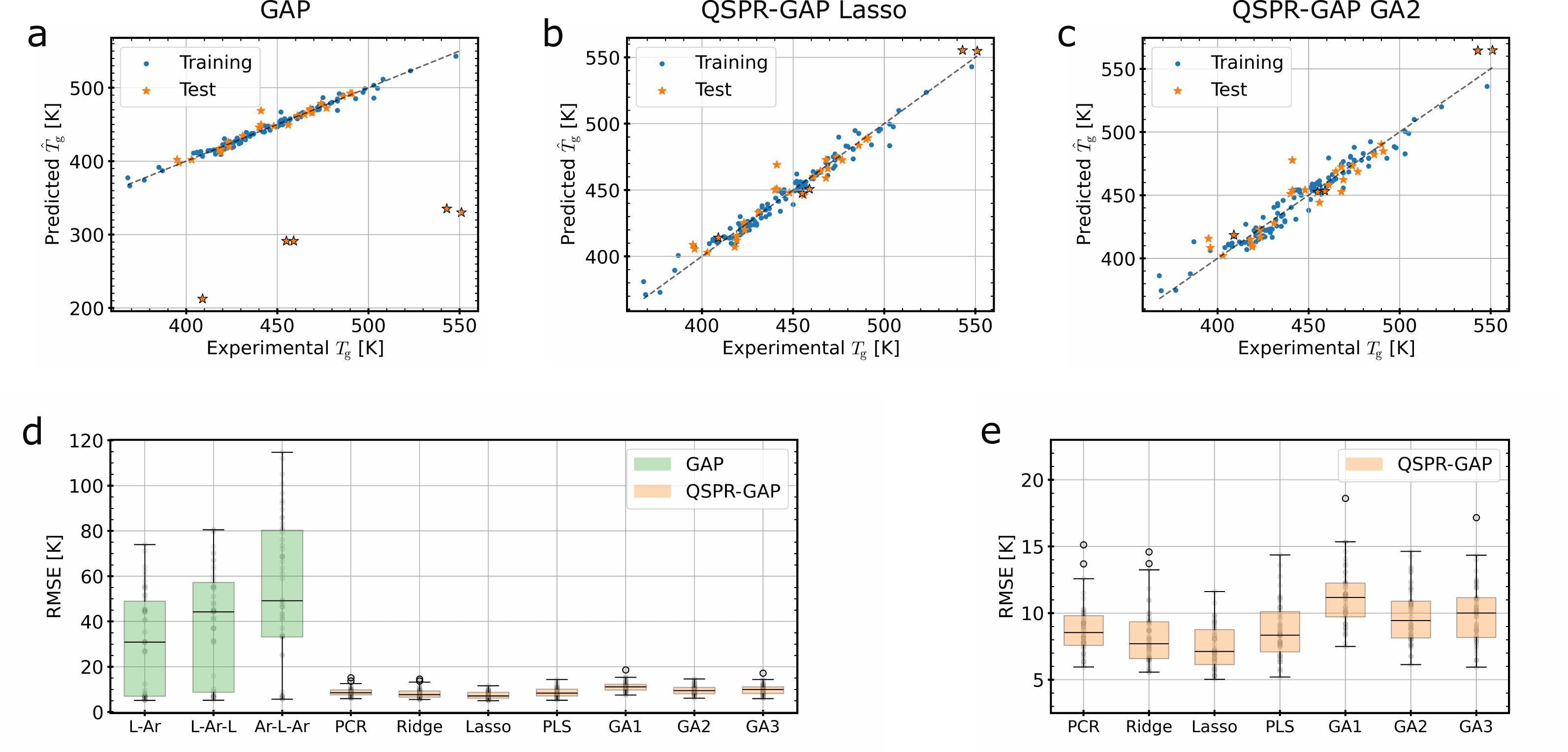}
\end{center}
\caption{\textbf{Comparison between the QSPR-GAP and the GAP model.} \textbf{a-c}~ Example for a randomly chosen training-test split where five polymers in the test set contain at least one out-of-sample fragment: \textbf{a}~GAP model, \textbf{b}~QSPR-GAP Lasso model, \textbf{c}~QSPR-GAP Genetic Algorithm with $\mG=2$ descriptors combined with OLS regression. \textbf{d}~Box and whisker plots summarising the distribution of root mean-squared error (RMSE) during external validation. The orange and green boxes represent the interquartile range (IQR), covering the second and third quartiles (Q2 and Q3) of the results data, with the black line indicating the median. Whiskers extend to the most extreme points within 1.5 times the IQR from Q1 and Q3, while outliers appear beyond this range.
\textbf{e}~The same QSPR-GAP model results as in \textbf{d} (expanded). The polymers containing out-of-sample fragments (outliers) are shown in \textbf{a-c} as orange stars with a black outline. }
\label{fig:data_plots}
\end{figure*}

\subsection{GAP and QSPR-GAP approaches}

The predicted glass transition temperature for the $a$th polymer $\Tgahat$ is represented as a molar-mass weighted average of the estimated $\Tgval$-contribution $\hat{\beta_i}$ from each $i$th fragment,
\begin{equation}
\Tgahat = \frac{\sum_{i=1}^p {X}_{ai} M_{i} \hat{\beta}_{i}}{\sum_{i=1}^p {X}_{ai} M_{i}}\equiv\sum_{i=1}^p \bar{X}_{ai} \hat{\beta}_i,
\label{eq:fitted_Tg}
\end{equation}
where $i$ indexes the fragments (as labelled in Fig.~\ref{fig:Flow_diagram}-B3), $M_i$ is the molar mass of the $i$th fragment, and thus, $\bar{X}_{ai}$ is the mass-weighted composition of fragment $i$ in polymer $a$. The polymer $\Tgval$ is thus modelled by its composition-weighted constituent fragment contributions, $\beta_i$, where $\beta_i$ corresponds to $\Tgval$ of a long-chain homopolymer composed entirely of the $i$th fragment. Since $\beta_i$ is unknown, it is estimated; note that we denote an estimated (or predicted) value by a hat~$\hat{}$.

We estimate $\beta_i$ in two different ways: (i) as a benchmark, we use a GAP approach based simply on the identity of the fragment; or (ii) a novel combined QSPR-GAP approach based on the molecular features of each fragment encoded in the descriptors. In the GAP approach, 
the count matrix $\tensor{X}$ is molar mass normalised (Eq.~\ref{eq:molar_mass_averaged_X}) giving the composition matrix $\X$ with elements $\bar{X}_{ai}$, and $\beta_i$ is estimated from the experimentally available $\Tgval$ values by Ordinary Least Squares (OLS) regression against $\X$ (Eq.~\ref{eq:GAP_estimator}).

In the QSPR-GAP approach, the key distinction from the GAP method lies in the parameterisation of 
$\beta_i$ (and consequently, $\Tgval$) by a set of molecular descriptors that encode the structure of each fragment (see Methods for a detailed description of how  $\beta_i$ is estimated). The $\Tgval$-contribution of fragment $i$ is expressed in terms of the values of the molecular descriptors $D_{i\mu}$, according to
\begin{equation}
\beta_i = \gamma_0 + \sum_{\mu=1}^m D_{i \mu} \gamma_{\mu}.
\label{eq:beta_estimation_qspr}
\end{equation}
Here, the regression coefficient $\gamma_{\mu}$ parameterises the influence of molecular descriptor $\mu$ on $\Tgval$, and $\gamma_0$ is a constant, both of which are estimated by the regression methods explained below. 
Since the inputs of Eq.~\eqref{eq:beta_estimation_qspr} are physical molecular descriptors rather than occurrences of a given fragment, the QSPR-GAP model can also be used to predict the $\Tgval$-value of polymers that contain a $j$th fragment that does not exist within the data sample. 

\subsection{Regression Methods}\label{section:Models}

Our data set of $n=146$ polymers with corresponding $\Tgval$ values was divided into $p=30$ unique $L\-Ar\-L$ fragments, and as a benchmark, a GAP analysis was performed using OLS to estimate the $\Tgval$ contributions $\beta_i$  of fragments $i=1,\ldots,p$. For the  QSPR-GAP analysis, in turn, the information about the $p=30$ fragments was encoded into $m=213$ molecular descriptors and four linear regression methods were used to determine $\hat{\gamma}_0$ and $\hat{\gamma}_\mu$ (for each descriptor $\mu=1,\ldots,m$): Principal Component Regression (PCR), Ridge regression, Lasso regression \cite{Lasso1996Tibshirani} and Partial Least Squares (PLS) regression \cite{PLS2000} (see SI Sec.~S-II for a brief discussion of each). These regression methods were chosen due to their robustness against overfitting, which would otherwise occur since the number of fit parameters in Eq.~\eqref{eq:beta_estimation_qspr} ($m+1$) exceeds the number of observed data points (\textit{i.e.} the $n$ polymers). The regression methods also account for the multicollinearity among the molecular descriptors (see Fig.~S5~and~S6; SI) by penalising the size of the estimated coefficients $\hat{\gamma}_{\mu}$, resulting in many fewer `effective' regression coefficients.

As an alternative implementation of QSPR-GAP, a genetic algorithm (GA) was applied to select the subset of $\mG$ descriptors (out of all $m=213$ descriptors) that best predict $\Tgval$ by linear regression (see SI Sec.~S-II for more details). Ten GA models were investigated, here termed ``QSPR-GAP GA$\mG$'' ($\mG=1,\ldots,10$), each resulting in different estimates for coefficients $\gamma_0$ and $\gamma_{\mu=1\ldots\mG}$, for the $\mG$ descriptors chosen.

\subsection{Performance of the QSPR-GAP model}

To assess how well a model generalises to new (or unseen) data it is essential to perform an external validation. Often, an external validation is performed on a reserved test set used only for this purpose, while model selection and/or tuning is performed on the training set (the remaining part of the data set) during an internal validation \cite{Pilania2019Machine-Learning-BasedCopolymers, Alesadi2022MachineStructure, Mattioni2002PredictionNetworks, ParandekarModelingRelationships2015}. A drawback in selecting a dedicated test set is possible selection bias, \textit{i.e.} bias due to random fluctuations in smaller data sets. To avoid this, we iteratively select different test sets such that all data points are eventually used in a test set.
Full details of the external and internal validations are outlined in SI Sec.~S-III.

Briefly, the external validation was performed using a repeated five-fold cross validation (5 fold-CV) where the full data set was shuffled randomly and subsequently partitioned into five distinct subsets. A test set was iteratively selected from the five subsets and in each iteration the remaining four subsets were combined into a single training set. An internal validation was performed on the training set at each iteration, in order to 
tune the model `hyperparameters' (\textit{e.g.}, the number of principal components in PCR and PLS, or the degree of shrinkage in Ridge and Lasso; see SI Sec.~S-II). This procedure was repeated 10 times, leading to 50 different training-test splits, each with a unique combination of polymers. One important aim of this procedure is to ensure that many test sets contain polymers with fragment IDs absent from the training set, which enables efficient probing of the robustness of our proposed QSPR-GAP approach. Such out-of-training set fragment occurrences, in the following referred to as `out-of-sample fragments', were identified a total of 34 times for the 50 different training-test splits (Fig.~S14; SI).

To demonstrate the effectiveness of our proposed QSPR-GAP method compared to the (benchmark) GAP method, Figs.~\ref{fig:data_plots}a-c display the predicted $\hat{\Tgval}$ versus the experimental $\Tgval$ for a representative training-test split across three different models, as representative examples. In these figures, blue circles represent the training set data, while orange stars indicate the test set data. We include results for the GAP model in panel (a), the QSPR-GAP Lasso model in panel (b), and the QSPR-GAP GA2 model in panel (c). For this particular partition (training-test split) of the data, the fragments with $i=8$, $i=20$, $i=22$ and $i=30$ (as labelled in Fig.~\ref{fig:Flow_diagram}-B3) do not exist in the training set. For the GAP model, these four out-of-sample fragments manifest as five clear outlier polymers, which the model can not handle (outlined orange stars in Fig.~\ref{fig:data_plots}a). Since these fragments are absent in the training set, the corresponding $\hat{\beta}_i$ values are not known and we thus set $\hat{\beta}_{8} = \hat{\beta}_{20} = \hat{\beta}_{22} = \hat{\beta}_{30} = 0$, leading to the outlier polymers; better predictions could be obtained by making ad hoc assumptions about the out-of-sample fragments
(for example, assuming an average value of the estimated in-sample $\beta_i$ contributions).

The advantages of the QSPR-GAP approach become obvious when comparing results from the same data partitions involving the same out-of-sample fragments. The QSPR-GAP Lasso model in Fig.~\ref{fig:data_plots}b and QSPR-GAP GA2 model in Fig.~\ref{fig:data_plots}c demonstrate a significantly more robust prediction than the GAP model for the five outlier polymers. 
The root-mean-squared error (RMSE) from the full external validation, \textit{i.e.} the results from the full 50 training-test splits, is presented in Fig.~\ref{fig:data_plots}d for all investigated GAP and QSPR-GAP models; with more detail shown in Fig.~\ref{fig:data_plots}e for the QSPR-GAP models. For the GAP model, in addition to the $L\-Ar\-L$ fragment definition, we also investigated the definitions $L\-Ar$ and $Ar\-L\-Ar$.
From 50 splits, out-of-sample fragments are found 30 times for the $L\-Ar$, 34 times for the $L\-Ar\-L$, and 40 times for the $Ar\-L\-Ar$ fragment choice. The increase in  out-of-sample fragments grows with the number of available combinations of $L$ and $Ar$ groups (Fig.~\ref{fig:Flow_diagram}A). More frequently occurring out-of-sample fragments, and the resulting outlier polymers, increase the RMSE in the GAP approach, from $\sim$30\,K to $\sim$45\,K to $\sim$50\,K (Fig.~\ref{fig:data_plots}d) for the three fragment definitions $L\-Ar$, $L\-Ar\-L$, and $Ar\-L\-Ar$, respectively.

Our QSPR-GAP approach demonstrates much better robustness against out-of-sample fragments than the GAP-based approach (Figs.~\ref{fig:data_plots}d-e). All seven QSPR-GAP models (PCR, Ridge, Lasso, PLS, GA1, GA2, and GA3) show a median RMSE ranging from 8-11\,K, compared with 30-50\,K for the GAP models. Moreover, the QSPR-GAP models maintain the robustness depicted in the single training-test example (Fig.~\ref{fig:data_plots}b-c) for all 50 splits. As shown in Fig.~\ref{fig:data_plots}e, even though all seven investigated QSPR-GAP models perform similarly, the Lasso model is the most accurate (by a few degrees K), with a RMSE range of $\simeq5-12\text{\,K}$ (depending on the partitioning of training/test data), and a median RMSE of $\simeq8\text{\,K}$. We note that the GA2 model shows good predictive ability, even though it is based only on two descriptors ($\mG=2$), thus requiring only three fit parameters (see Eq.~\ref{eq:beta_estimation_qspr}); with  a predictive RMSE range of $\simeq6-15\textrm{\,K}$. Since the predictive ability (characterised by the RMSE) for the GAP models is significantly affected by the outlier polymers caused by out-of sample fragments, we have also compared models for which all outliers were removed (Fig.~S15; SI). We find that the predictions of the QSPR-GAP models are slightly improved, as exemplified by a RMSE~$\simeq5\textrm{-}9\,\textrm{K}$ for the Lasso method, whereas the GAP models demonstrate a highly improved RMSE~$\simeq5\textrm{-}8\,\textrm{K}$.

\begin{figure*}[htbp]
\begin{center}
\includegraphics[width=.85\textwidth]{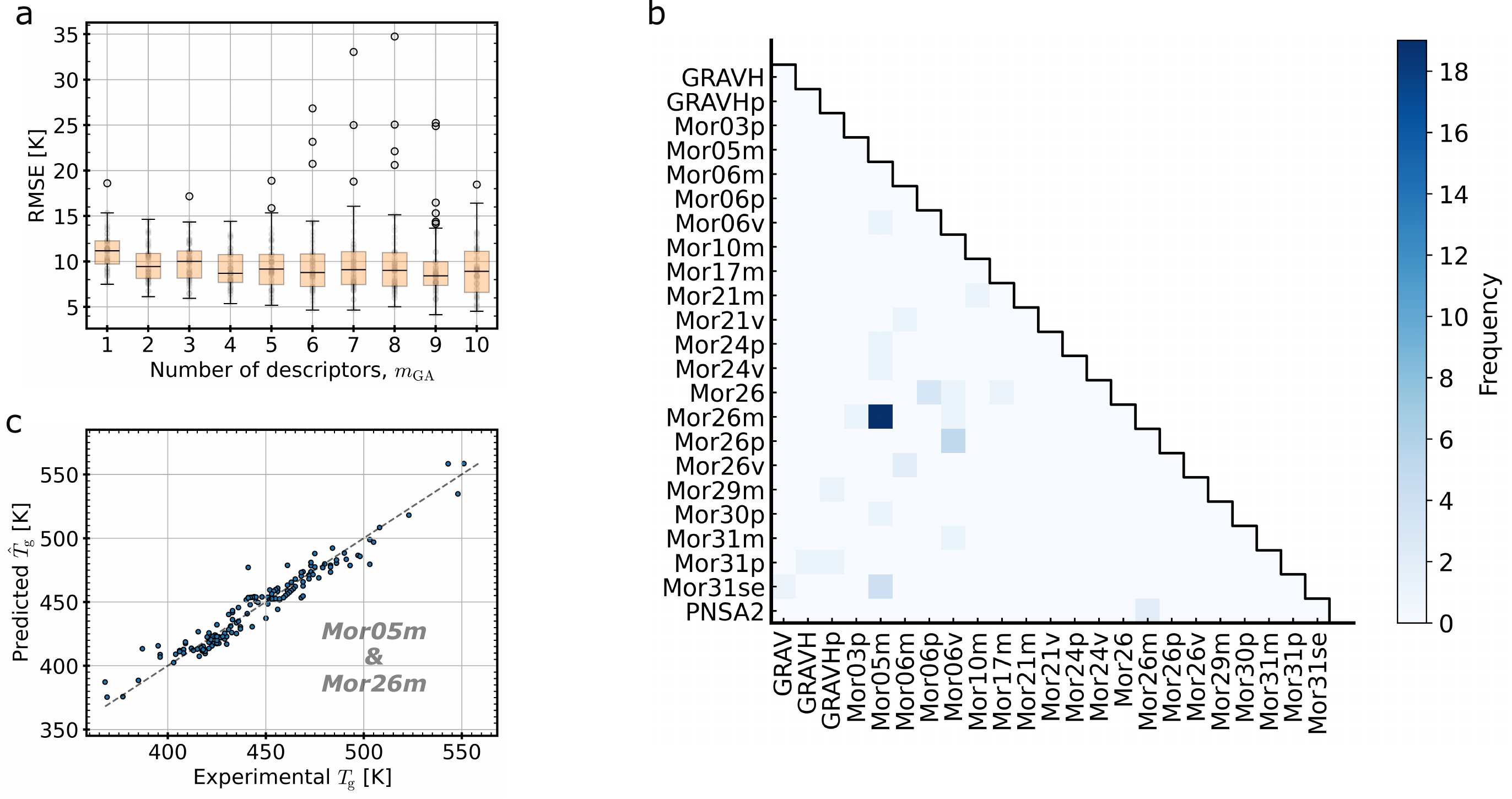}
\end{center}
\caption{\textbf{Analysis of descriptors.} \textbf{a} Box and Whisker
plots (see caption of \protect{Fig.~\ref{fig:data_plots}} for an explanation) 
summarise the distribution of the Root Mean Squared Error (RMSE) for ten Genetic Algorithm (GA) models, consisting of a GA-based selection of $\mG=1, \mG=2,\ldots\,\mG=10$ optimal descriptors (from a pool of 213), 
followed by OLS regression. The RMSE is determined from a 5-fold cross validation repeated ten times, resulting in 50 training-test splits. \textbf{b}~GA results for $\mG=2$: from 
the complete set of two-descriptor combinations we show all pairs selected at least once. As shown, the descriptor pair \texttt{Mor05m} and \texttt{Mor26m} was selected 19 times for the 50 training-test splits. \textbf{c}~Predicted (in-sample) vs actual $\Tgval$ values from an OLS regression on the full dataset based only on the two descriptors \texttt{Mor05m} and \texttt{Mor26m}.}
\label{fig:GA_analysis}
\end{figure*}

\section{Descriptor analysis}
\subsection{Identifying the most important descriptors}

The key feature of the genetic algorithm models (QSPR-GAP GA$\mG$) is that they explicitly select a subset of ($\mG=1,\ldots,10$) descriptors that best predict $\Tgval$. In testing the performance of the GA models, the same external validation consisting of a 5-fold CV (repeated 10 times) is used, resulting in a total of 50 independent out-of-sample predictions, \textit{i.e.} 50 different training-test splits (SI, Sec.~S-III).

The RMSE results for the ten investigated QSPR-GAP GA$\mG$ models are shown in Fig.~\ref{fig:GA_analysis}a. Remarkably, only a few descriptors are needed for good prediction. The most significant improvement in the predictive accuracy occurs between $\mG=1$ and $\mG=2$, while for $\mG>2$ there is no significant improvement. Hence, only two descriptors are necessary (for this data sample) to predict $\Tgval$ with an RMSE $\simeq 6-15\,\textrm{K}$. Thus, in the following we restrict our attention to the two-descriptor model, QSPR-GAP GA2. 

Fig.~\ref{fig:GA_analysis}b illustrates the frequency by which the GA selected different pairs of descriptors. It is clear that one pair stands out: \texttt{Mor05m} and \texttt{Mor26m}. From the 22,578 possible pairs arising from 213 descriptors, this pair was selected 19 out of 50 times (Fig.~\ref{fig:GA_analysis}b), from the 50 random data partitions.
The descriptors \texttt{Mor05m} and \texttt{Mor26m} belong to the 3D-MoRSE family (Molecular Representation of Structures based on Electronic diffraction) \cite{Schuur1996TheActivity, Schuur1997InfraredRepresentation, Devinyak20143D-MoRSEExplained}, which describes the 3D structure of a given fragment (or molecule) by a `form factor' based on atom-to-atom pair distances,
\begin{equation}
    I(q) = \sum_{l=1}^{N-1} \sum_{k=l+1}^{N} A_k A_l \frac{\sin(q \, r_{kl})}{q \, r_{kl}},
    \label{eq:3D_MoRSE}
\end{equation}
where $k$ and $l$ label specific atoms, $N$ is the number of atoms in the fragment, $A_k$ and $A_l$ are  weighting factors for the atoms $k$ and $l$, $q$ is the `scattering' wave vector, and $r_{kl}$ is the Euclidean distance between atoms $k$ and $l$.

Our descriptor set ($m=213$ descriptors calculated using \texttt{Mordred})  contains 160 3D-MoRSE descriptors, characterised by 32 different $q$ values: $0$ (\texttt{Mor01}), $1\,\textrm{\AA}^{-1}$ (\texttt{Mor02}), $2\,\textrm{\AA}^{-1}$ (\texttt{Mor03}), \ldots, $31\,\textrm{\AA}^{-1}$ (\texttt{Mor32})) and five different weighting schemes for $A_k,A_l$: unweighted ($A_k=A_l=1$), atomic mass (\texttt{MorXXm}), van der Waals atomic volume (\texttt{MorXXv}), Sanderson electronegativity (\texttt{MorXXse}) \cite{SandersonElectronegativity}
and polarisability (\texttt{MorXXp}); all weighting schemes are scaled by their value for carbon. The MoRSE descriptors \texttt{Mor05m} and \texttt{Mor26m} correspond to $q=4\,\textrm{\AA}^{-1}$ and $q=25\,\textrm{\AA}^{-1}$, where $A_i$ is the ratio of the mass of atom $i$ to the mass of carbon.

The $\Tgval$ contribution $\hat{\beta}_i$ for fragment $i$ can thus be accurately estimated by
\begin{equation}
    \hat{\beta}_i = \hat{\gamma}_0 + \hat{\gamma}_1 I_i(4) + \hat{\gamma}_2 I_i(25),
    \label{eq:beta_from_Mor05m_Mor26m}
\end{equation}
where $I_i(4)$ and $I_i(25)$ are calculated from the set of atoms in fragment $i$ and $\hat{\gamma}_{\mu=0,1,2}$ are the regression coefficients, estimated with an OLS regression applied to the full data sample of 146 polymers (no training-test splits); these estimates are provided in Table~\ref{table:regression_coefs}.
The predicted against experimental $\Tgval$ results are shown in Fig.~\ref{fig:GA_analysis}c, with an in-sample RMSE~$\simeq8\,\textrm{K}$.

\begin{table}[H]
    \centering
    \rowcolors{2}{white}{gray!25}
    \begin{tabular}{cccc}   
        \hline\hline
        \textrm{$\mu$} & \textrm{Descriptor} & \textrm{$\hat{\gamma}_\mu$ [\textrm{K}] } & \textrm{CI (95\%) L/U [K]} \\
        \hline
        \hline
        0 & \textit{--} & 298 & 286/310 \\
        1 & \texttt{Mor05m} & -58  & -67/-50 \\
        2 & \texttt{Mor26m} & -198  & -239/-157 \\
        \hline\hline
    \end{tabular}

    \caption{\textbf{Regression coefficients} from the best two-descriptor Genetic Algorithm model, as estimated from the full dataset of 146 polymers (OLS). Upper and lower $95\%$ confidence intervals (CI) are presented for the estimated coefficients $\gamma_0$ and $\gamma_\mu$. For assumptions and diagnostics of distributions, see SI, Sec. S-IV.}
    \label{table:regression_coefs}
\end{table}

\subsection{Atomic level $\Tgval$ contributions}
\begin{figure*}[htbp]
\begin{center}
\includegraphics[width=.85\textwidth]{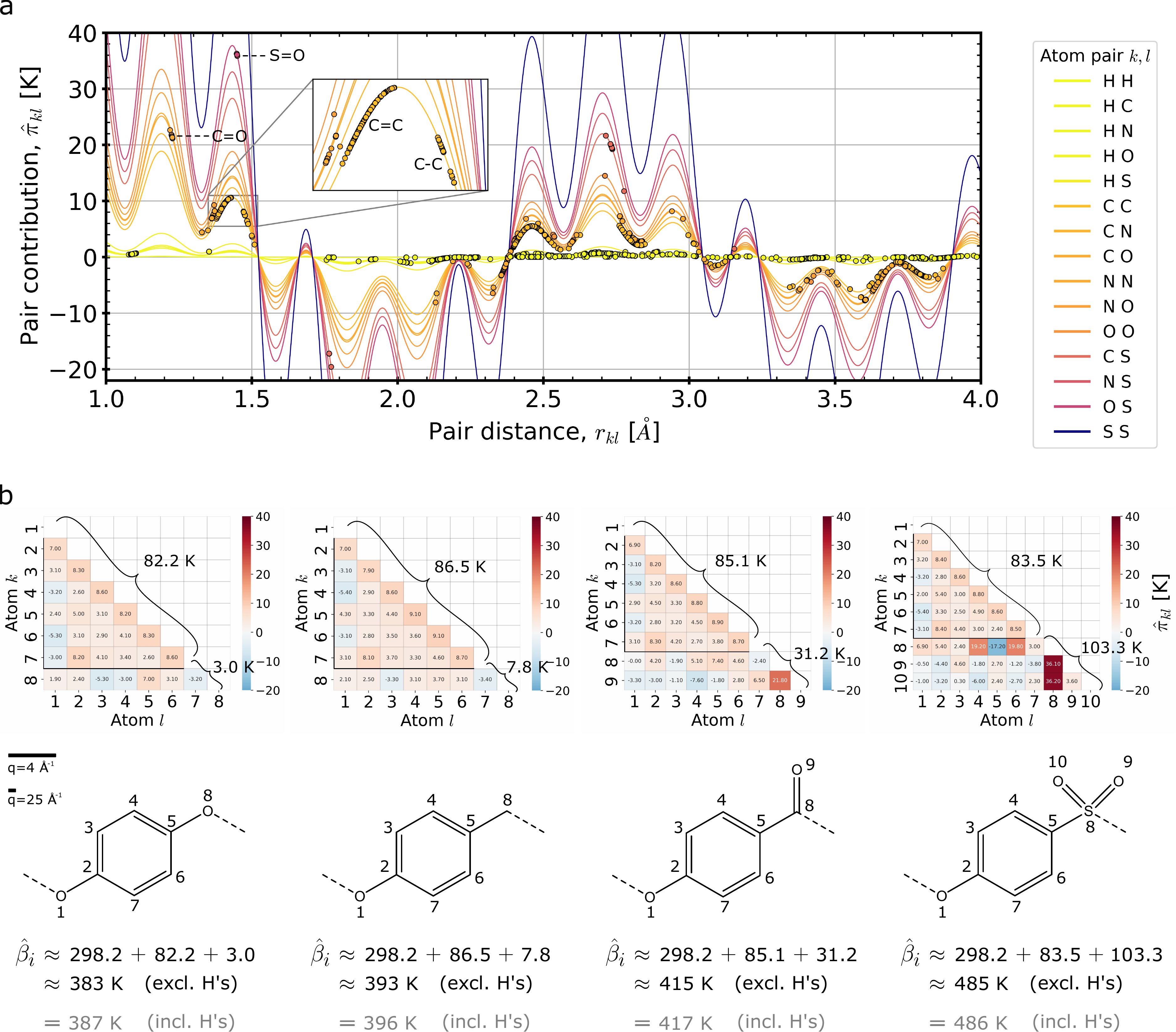}
\end{center}
\caption{\textbf{Estimated inter-atomic pair $\Tgval$ contributions.} \textbf{a} The estimated function $\hat{\pi}_{kl}$ of pairwise atomic $\Tgval$ contributions, as expressed in Eq.~\eqref{eq:pair_contributions}; $\hat{\pi}_{kl}$ is a function of the pair distance $r_{kl}$ between $k$ and $l$ atoms, and the product $M_kM_l$, where the latter is illustrated by the colour gradient. The lines represent the estimated function $\hat{\pi}_{kl}(r_{kl}, M_kM_l)$, and the solid points represent the function evaluated for the specific atom pairs in the data set. The inset is a magnification of the range containing contributions from different carbon-carbon single and double bonds, whose bond lengths (when energy-minimised) vary slightly depending on the specific fragment. \textbf{b}~Four example fragment IDs that share the same structure (in atoms 1-7) but vary by a single functional group (in atoms 8+). The pairwise contributions $\hat{\pi}_{kl}$ are shown as coloured tiles with their values shown inside each tile. The scale bars show the length-scales (1.6\textrm{\AA} and 0.3\textrm{\AA}) corresponding to $q=4\,\textrm{\AA}^{-1}, 25\,\textrm{\AA}^{-1}$, relative to the size of the (planar) structures. In the sum over pair contributions (Eq.~\ref{eq:fragment_Tg_contribution_sum}) atoms $l = 1, \ldots, 6$ and $k=2, \ldots, 7$ where $l<k$ (indicated by the partition) correspond to the same structure and yield nearly the same contribution $\Delta \hat{\beta}_i$ $\simeq82.2-85.5\,\textrm{K}$ for all four fragments. The remaining atom pairs set the fragments apart in their summed contribution to $\hat{\beta}_i$, which vary from $\Delta \hat{\beta}_i\simeq3.0-103.3\,\textrm{K}$. Atomic pair contributions including hydrogen atoms have been ignored from the plot since they show only small contributions to $\hat{\beta}_i$ (according to Eq.~\ref{eq:pair_contributions}).}
\label{fig:MoRSE_analysis}
\end{figure*}

Using Eqs.~\eqref{eq:3D_MoRSE} and \eqref{eq:beta_from_Mor05m_Mor26m} we can express the estimated $\Tgval$ contribution of fragment $i$ as a sum over atomic pair contributions
\begin{equation}
    \hat{\beta}_i = \hat{\gamma}_0 + \sum_{l=1}^{N_i-1} \sum_{k=l+1}^{N_i} \hat{\pi}_{kl},
    \label{eq:fragment_Tg_contribution_sum}
\end{equation}
where $\hat{\pi}_{kl}$ denotes the (estimated) $\Tgval$ contribution given by the pair of atoms $k$ and $l$, expressed in terms of the two descriptors \texttt{Mor05m} and \texttt{Mor26m}
\begin{equation}
    \hat{\pi}_{kl} = \hat{\gamma}_1 M_k M_l \frac{\sin(4 r_{kl})}{4 r_{kl}} + \hat{\gamma}_2 M_k M_l \frac{\sin(25 r_{kl})}{25 r_{kl}};
    \label{eq:pair_contributions}
\end{equation}
$M_k$ is the mass of atom $k$ (divided by the mass of carbon) and $r_{kl}$ is the distance between atom pair $k,l$.
To determine the fragment contribution $\hat{\beta}_i$, the $\hat{\pi}_{kl}$ contributions are summed over the total number of atoms $N_i$ in the $i$th fragment according to Eq.~\eqref{eq:fragment_Tg_contribution_sum}.

The function $\hat{\pi}_{kl}(r_{kl}, M_kM_l)$ is shown in Fig.~\ref{fig:MoRSE_analysis}a and at constant $r_{kl}$, $\hat{\pi}_{kl}$ is a linear function of the product $M_kM_l$ of atomic masses, as illustrated by the colour gradient, and at constant $M_kM_l$, $\hat{\pi}_{kl}$ is an oscillating function of the pair distance. As $r_{kl}$ increases, the pair contributions $\hat{\pi}_{kl}$ become less relevant to the overall $\hat{\beta}_i$, with  negative and positive contributions cancelling out in the summation of Eq.~\eqref{eq:fragment_Tg_contribution_sum}. Thus, the most important contributions lie within the range $r_{kl}\simeq1.2-1.5\textrm{\,\AA}$.

In Fig.~\ref{fig:MoRSE_analysis}b we show four $L_1\-Ar\-L_2$ fragments for which only the linker $L_2$ differs. As discussed above, the structure of each fragment is energy-minimised (using MMFF) and is represented by a unique set of $r_{kl}$, $M_k$ and $M_l$ values. For each of the four fragments, the function $\hat{\pi}_{kl}(r_{kl},M_kM_l)$ is evaluated for all atomic pairs and the results are provided in the tiles (also illustrated by the corresponding heat map); each tile represents the pairwise atomic $\Tgval$ contribution from atoms $k$ and $l$. The overall fragment $\Tgval$ contribution, $\hat{\beta}_i$, results from a sum over all atomic pair contributions, $\hat{\pi}_{kl}$, and the constant $\hat{\gamma}_0$.

Since hydrogen-containing pairs show very small $\Tgval$ contributions (as shown in Fig.~\ref{fig:MoRSE_analysis}a), they are omitted for clarity in Fig.~\ref{fig:MoRSE_analysis}b. The contributions from atoms 1 through 7 are very similar because these atoms represent the same molecular structure motif (ether-linked phenyl). Hence, the sum over $\hat{\pi}_{kl}$ for atoms $k,l = 1, \ldots, 7$ gives contributions of $\Delta\hat{\beta}_i$ = 82.2~K, 86.5~K, 85.1~K, and 83.5~K, respectively for the four fragments, as shown in Fig.~\ref{fig:MoRSE_analysis}b. The slight differences between these values arise when minimising the energy, because the inter-atomic distances $r_{kl}$ are influenced by the atoms in linker $L_2$. The differences in the linker $L_2$ (atoms 8+) leads to significant differences in $\Tgval$, with contributions of $\Delta \hat{\beta}_i$ = 3.0~K, 7.8~K, 31.2~K, and 103.2~K for the four structures. The total $\Tgval$ contributions for the four fragments are  $\hat{\beta}_i$ = 383~K, 393~K, 415~K, and 485~K, respectively, determined using Eq.~\eqref{eq:fragment_Tg_contribution_sum}, while excluding hydrogens from the sum.

We conclude that for PAEK polymers the $\Tgval$ contribution of each fragment ($\beta_i$) is very well approximated as a constant plus a sum over all atomic pair contributions ($\hat{\pi}_{kl}$), where the main contributions correspond to short atomic pair-distances of $1.2-1.5\,\textrm{~\AA}$.
The genetic algorithm identified two important 3D-MoRSE descriptors corresponding to length-scales ($\sim 2\pi/q$) of 0.3 and 1.6 $\textrm{\AA}$ (see the scale bars in Fig.~\ref{fig:MoRSE_analysis}b) and we anticipate these are PAEK-family specific quantities.
The example shown in Fig.~\ref{fig:MoRSE_analysis}b, demonstrates that the linker properties, such as their bulkiness, have the greatest impact on $\Tgval$. For the glass transition in general, the interplay between packing and chain flexibility \cite{Baker2022PRX,matsuoka1997entropy} suggests that three-body features, such as intramolecular angles, play an important role. In this study, such features are likely implicitly incorporated due to the chosen $L\-Ar\-L$ motif.

\section{Conclusion}

We present a new method for predicting $\Tgval$ from the monomer structure in polymers.
The method combines group contributions, or equivalently group additive properties (GAP), with a quantitative structure property relationship (QSPR) approach.
The GAP method assumes $\Tgval$ can be expressed by a composition-weighted average over $\Tgval$ contributions from sub-monomer motifs (fragments), and our QSPR-GAP model uses molecular descriptors to relate the physical properties of these fragments to their GAP-like $\Tgval$ contributions.
We apply this model to a dataset of 146 linear poly(aryl ether ketone) (PAEK) homo- and co-polymers, resulting in a median root-mean-square error of 8 K (out-of-sample).

Descriptor calculations for sub-monomer fragments are significantly faster than in traditional QSPR approaches based on monomers (or oligomers). In addition, the QSPR-GAP method yields accurate predictions for polymers containing fragments outside the data sample, which resolves the main limitation of the GAP approach. Using a genetic algorithm, we show that only two  molecular descriptors (from a pool of 213) are necessary to predict $\Tgval$ with an RMSE $\simeq 6-15\,\textrm{K}$. Moreover, we identify a direct mapping between $\Tgval$ and the monomer structure through pairwise atomic contributions.

This work offers an accurate, accessible, and broadly applicable predictive model, suitable for small data sets and deployment on a standard laptop (within minutes). The QSPR-GAP method is transferable to other classes of polymers, both synthetic and natural (\textit{e.g.} conjugated or bio-polymers), and to physical behaviour beyond the glass transition, such as mechanical, optical, or transport properties.

\section{Methods}

The number of occurrences of fragment $i$ in polymer $a$ is
\begin{equation}
(\tensor{X})_{ai}\equiv X_{ai}
\end{equation}
where $\tensor{X}$ is an $n \times p$ dimensional count matrix, with $n$ rows representing the full set of polymer IDs, and $p$ columns representing the full set of unique fragment IDs. We normalise $\tensor{X}$ by the molar mass of the repeating unit, resulting in the mass-weighted composition matrix
\begin{equation}
    \bar{\tensor{X}} = (\text{diag}[{\tensor{X}\vec{M}}])^{-1}\tensor{X} (\text{diag}[\vec{M}]),
    \label{eq:molar_mass_averaged_X}
\end{equation} 
where $\vec{M}\in{\mathbb{R}}^p$ is a $p$-vector that enumerates the fragment molar masses. Note that the molar mass $M_i$ of an $L\-Ar\-L$ fragment is the molar mass of half of each $L$ group and the full $Ar$ group: $M_i=M_{L_{i1}}/2 + M_{Ar_i} + M_{L_{i2}}/2$. Since the same $L$ groups are counted twice when building a repeat unit structure from a given set of fragment IDs, the product $\tensor{X}\vec{M}$ encompasses the molar mass of the repeating unit for all polymers in the dataset (or correspondingly the molar mass of a copolymer's repeating unit, averaged over its comonomer mass fractions).

\subsection{GAP model}

We use ordinary least squares (OLS) to estimate the $p$ coefficients $\hat{\beta_i}$, \textit{i.e.} the $p$-vector $\hat{\b} \in \mathbb{R}^p$, by minimising the residual sum of squares:

\begin{equation}
    \hat{\b} = \underset{\b}{\arg\min} \left\{ \sum_{a=1}^n(\Tga-f((\tensor{\bar{X}})_a))^2 \right\},
    \label{eq:Least_squares_GAP}
\end{equation}
where the linear fitting function $f$ that approximates  $\Tga$ (of the $a$th polymer in the training data) is given by $f((\tensor{\bar{X}})_a) = \sum_{i=1}^p \bar{X}_{ai}\beta_i\equiv(\X \b)_a $. The solution, \textit{i.e.} the least squares estimate of $\b$, is given by
\begin{equation}
\hat{\b} = (\X^\top \X)^{-1} \X^\top \Tgvec,
\label{eq:GAP_estimator}
\end{equation}
where the $n$-vector $\Tgvec\in\mathbb{R}^n$ contains all $n$ $\Tga$ values in the training sample. Note that each estimated coefficient $\hat{\beta}_i$  corresponds to the predicted glass transition temperature of a polymer solely comprising fragment $i$ as its repeating monomer.

An out-of-sample prediction of the glass transition temperature for a polymer $b$ with fragment composition $\bar{X}_{bi}$ (given all $i=1, \ldots, p$) can now be determined as
\begin{equation}
    \Tgbhat = \sum_{i=1}^p \bar{X}_{bi} \hat{\beta}_i.
\end{equation}
Predictions of $\Tgval^{\,b}$ are restricted to polymers consisting of fragments whose contributions $\beta_i$ have already been estimated from Eq.~\eqref{eq:GAP_estimator}, meaning, GAP predictions are chemically constrained to polymers consisting of fragments within the set $\{1,\ldots,p\}$.

\subsection{QSPR-GAP model}

The QSPR-GAP model contains a \textit{descriptor matrix} $D_{i\mu}$, which encodes the chemical and physical properties of the $i$th fragment ID in terms of  $\mu=1,\ldots,m$ descriptor values,
\begin{align}
(\tensor{D})_{i\mu} &\equiv D_{i\mu}.
\end{align}
We then express each fragment $\Tgval$ contribution $\beta_i$ as a linear combination of the $m$ descriptors:
\begin{equation}
   \beta_i = \gamma_0 + \sum_{\mu=1}^m D_{i\mu} \gamma_\mu,  
\end{equation}
where the $(m+1)$-vector $\g \in \mathbb{R}^{(m+1)}$ contains the regression coefficients. The zeroth column index is included in the matrix $\D$ as $D_{i0}=1$ for all fragments $i = 1, \ldots, p$, to accommodate the constant term $\gamma_0$;  therefore $\D\in\mathbb{R}^{p\times(m+1)}$.

The methods used to estimate $\g$ include (1) Principal Component Regression, (2) Ridge regression, (3) Lasso regression, and (4) Partial Least Squares regression are discussed further in  the SI, Sec.~S-II. However, to illustrate the application of the QSPR-GAP model in its simplest form, we discuss the genetic algorithm (GA) model here.

The GA uses concepts analogous to evolution to select an optimal subset $(\mG\leq10)$ from a total of 213 potential descriptors. The descriptors chosen will have the greatest influence on $\Tgval$, and, once selected, are included in the function

\begin{subequations}
\label{eq:fitting_func_qspr_gap}
\begin{align}
   f((\tensor{\bar{X}})_a, \tensor{D}) &= \sum_{i=1}^p \bar{X}_{ai} \beta_i\\
   &= \gamma_0 + \sum_{i=1}^p \sum_{\mu=1}^{\mG} \bar{X}_{ai} D_{i\mu} \gamma_\mu
\end{align}
\end{subequations}
(since $\sum_{i=1}^p \bar{X}_{ai}=1$),
followed by a least squares minimisation

\begin{equation}
    \hat{\g} = \underset{\g}{\arg\min} \left\{ \sum_{a=1}^n(\Tga - f((\tensor{\bar{X}})_a, \tensor{D}))^2 \right\},
\end{equation}
yielding the solution
\begin{equation}
\hat{\g} = \left(\D^\top\X^\top\X\D\right)^{-1}\D^\top \X^\top \Tgvec.
\label{QSPR estimator}
\end{equation}

Once the coefficients $\g$ are estimated from the training data, the estimated $\Tgval$ contribution of any new fragment $j$ is given by 
\begin{equation}
\hat{\beta}_j = \hat{\gamma}_0 + \sum_{\mu=1}^{\mG} D_{j\mu} \hat{\gamma}_\mu,
\label{QSPR beta estimated}
\end{equation}
which may include fragments that are not in the original data sample.

An out-of-sample predicted glass transition temperature for a new polymer $b$ given fragment compositions $\bar{X}_{bj}$ (for all $j=1,\ldots,q)$ and molecular descriptors $D_{j\mu}$ (for all $\mu=1, \ldots, \mG)$, is
\begin{equation}
    \Tgbhat = \hat{\gamma}_0 + \sum_{j=1}^q \sum_{\mu=1}^{\mG} \bar{X}_{bj} D_{j\mu} \hat{\gamma}_\mu;
\end{equation}
noting that if $\{1,\ldots,p\}$ is the set of in-sample fragments, then $\{1,\ldots,q\}$ is the set of in-sample and out-of-sample fragments where $\{1,\ldots,p\} \subseteq \{1,\ldots,q\}$.

-------------
\section{Data availability}

Source data files are available at the University of Leeds Data Repository at https://doi.org/10.5518/1596. 

\section{Code availability}

The codes that support the findings of this study are available upon reasonable request from the corresponding author. 



%

\section{Acknowledgments}
The authors acknowledge financial support from the Engineering and Physical Sciences Research Council (EPSRC) funded Centre for Doctoral Training in Soft Matter and Functional Interfaces (grant EP/L015536/1) and from Victrex PLC. We thank Victrex PLC for the data they made available to conduct this study. PDO thanks the Ives foundation and Georgetown University for support. RJM thanks UKRI for funding via Future Leaders Fellowship (grant MR/W006391/1) and the University of Leeds for funding via a University Academic Fellowship.

\section{Author contributions}
S.B.C., P.D.O. and J.M. co-wrote the manuscript. S.B.C. performed all of the data analysis. J.M. and P.D.O. supervised the project. All
the authors reviewed and edited the manuscript and
contributed to useful discussions.

\section{Competing interests}

The authors declare no competing interests.

\end{document}


\title{Supplementary Information for: \\ 
A fast transferable method for predicting the glass transition temperature of polymers from chemical structure}

\author{Sebastian Brierley-Croft$^1$} 
	\author{Peter D.~Olmsted$^2$}
 	\author{Peter J.~Hine$^1$} 
    \author{Richard J.~Mandle$^1$} 
    \author{Adam Chaplin$^3$}
    \author{John Grasmeder$^3$} 
	\author{Johan Mattsson$^1$}
    \email{k.j.l.mattsson@leeds.ac.uk}
	\affiliation{$^1$School of Physics and Astronomy, University of Leeds, Leeds LS2\,9JT, United Kingdom}
	\affiliation{$^2$Department of Physics and Institute for Soft Matter Synthesis and Metrology, Georgetown University, Washington DC, 20057}
	\affiliation{$^3$Victrex PLC, Hillhouse International, Thornton Cleveleys, Lancashire FY5 4 QD, United Kingdom}
	\date{\today}
\maketitle

\tableofcontents
\newpage
\section{Data}
\subsection{The data set of glass transition temperatures}
\begin{figure}[H]
     \centering
     \includegraphics[width=.45\textwidth]{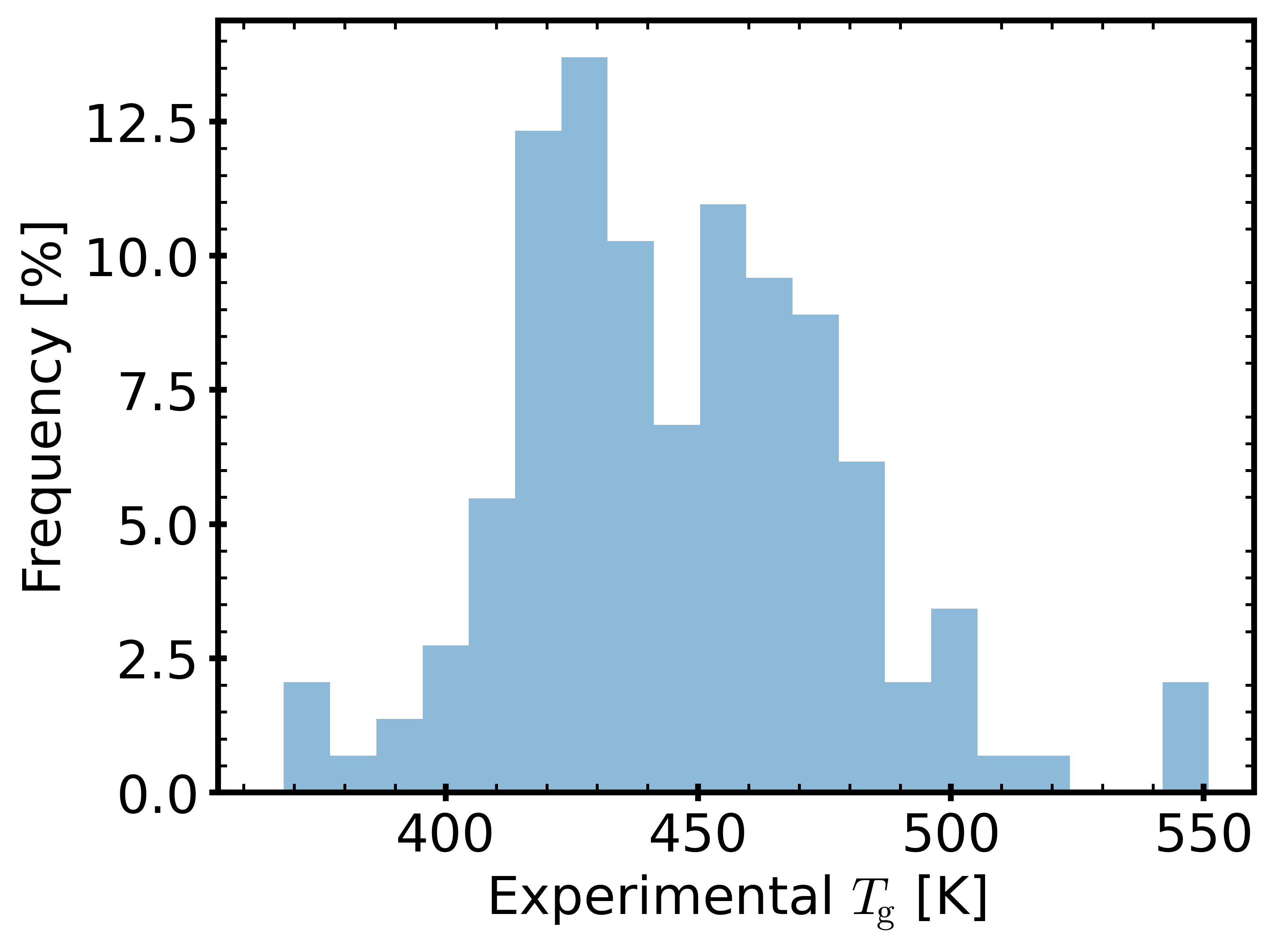}
     \caption{\textbf{Distribution of $\Tgval$ values in the dataset.}}
     \label{SI_fig:histogram}
\end{figure}

The data set contains glass transition temperatures ($\Tgval$) for a total of 77 homopolymers and 69 copolymers collected from an extensive list of references, and from measurements performed by the research and development team at Victrex.
The distribution of the $\Tgval$-values in the data set is shown in Fig.~\ref{SI_fig:histogram}.

\begin{figure}[tbh]
    \centering
    \includegraphics[width=0.75\textwidth]{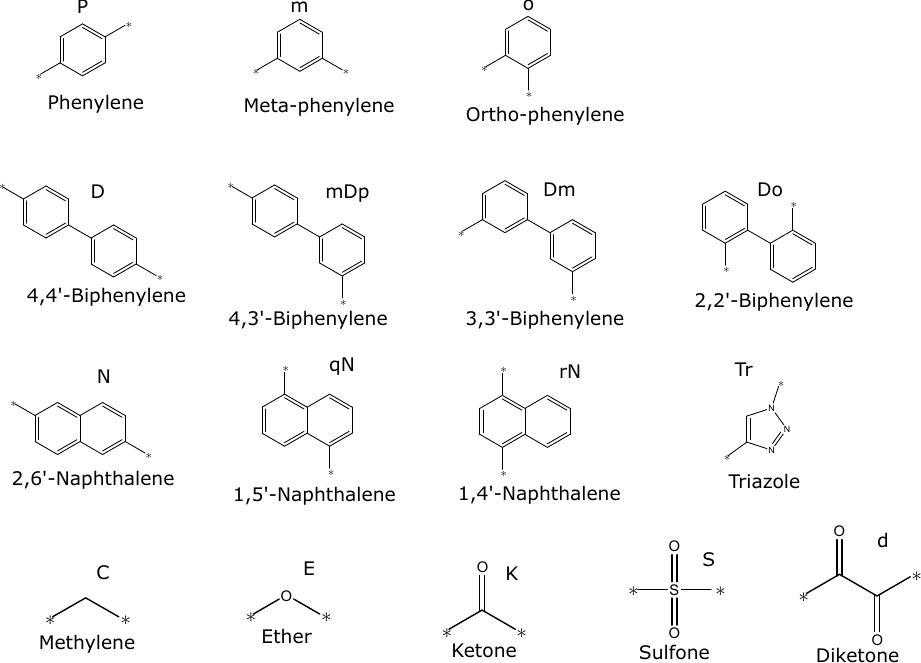}
    \caption{\textbf{Backbone Groups} that comprise the PAEK polymers in this data set. The first three rows are the rigid aryl groups, while the bottom row shows the flexible linkers.}
    \label{SI_fig:all_groups_SI}
\end{figure}
\begin{figure}[tbh]
    \centering
    \includegraphics[width=\textwidth]{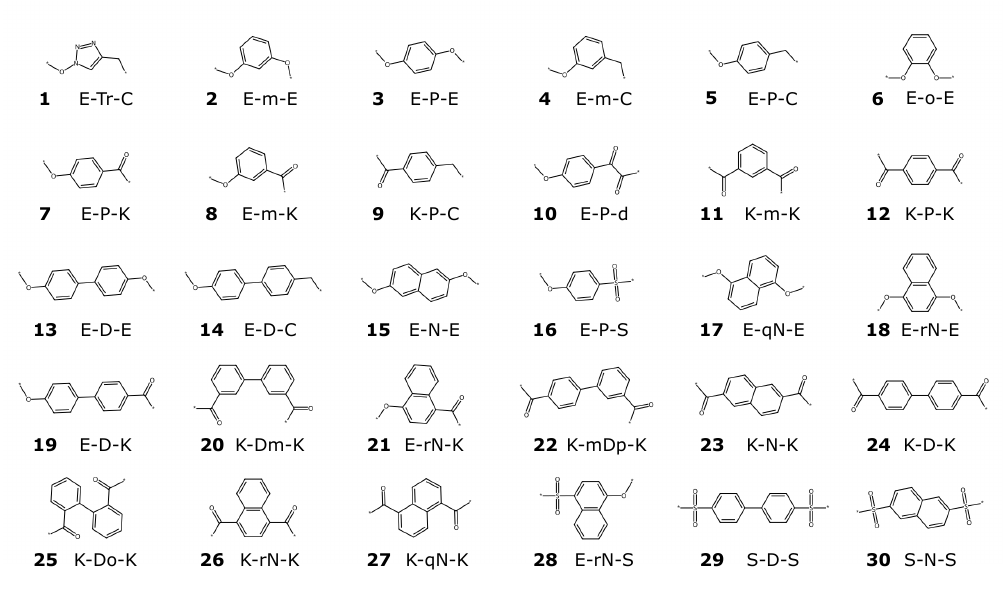}
    \caption{\textbf{All fragment structures} $i=1,\ldots,30$ present in the data set ($L\textrm{-}Ar\textrm{-}L$). The names of the functional groups follow Fig.~\ref{SI_fig:all_groups_SI}.}
    \label{SI_fig:all_fragments_SI}
\end{figure}

\subsection{3D-descriptor calculation}

The process involved in calculating the 213 3D-descriptors is as follows:
\begin{enumerate}
    \item The set of $p=30$ $L\-Ar\-L$ fragments were end-capped with hydrogen atoms, with the $i$th fragment ($i=1,\ldots,p$)  represented as a simplified molecular-input line-entry system (SMILES) string \cite{Weininger1988SMILESRules, Weininger1989SMILES.Notation}.
    \item Each SMILES string was converted into a \texttt{rdkit.Chem.rdchem.Mol} object using RDKit \cite{RDkit}, to create unoptimised 2D atomic coordinates for the $i$th fragment.
    \item A total of 500 initial 3D-conformations of each  fragment were generated using the RDKit function \\  \texttt{rdkit.AllChem.EmbedMultipleConfs(mol, numConfs=500, params=AllChem.ETKDG())}.
    \item The 3D-coordinates of the 500 conformations for each fragment were energy-optimised based on the Merck Molecular Force Field (MMFF), using the RDKit function \texttt{rdkit.AllChem.MMFFOptimizeMoleculeConfs(mol, maxIters=1000)}. This resulted in a distribution of optimised conformer energies from which the lowest energy conformer was chosen for each fragment.
    \item \texttt{Mordred} \cite{Moriwaki2018Mordred:Calculator} was used to calculate ($m=213$) 3D-descriptors $D_{i\mu}$ ($\mu=1,\ldots,m)$ of the lowest energy conformer for each of the ($p=30$) fragments.

\end{enumerate}

\begin{figure}[bth]
    \centering
    \begin{subfigure}[b]{0.49\textwidth}
        \centering
        \includegraphics[width=\textwidth]{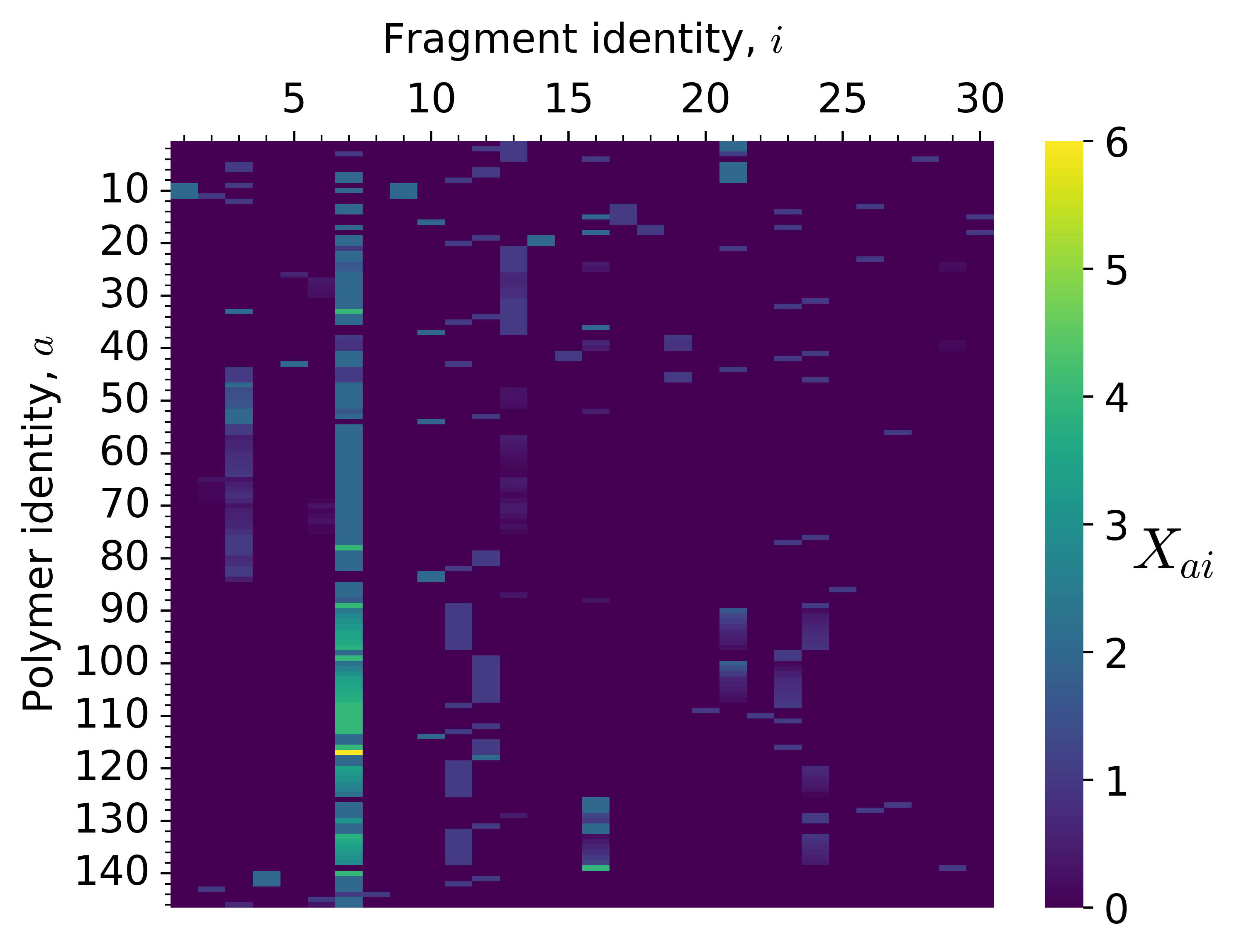}
        \caption{$\tensor{X}$}
    \end{subfigure}
    \hfill
    \begin{subfigure}[b]{0.49\textwidth}
        \centering
        \includegraphics[width=\textwidth]{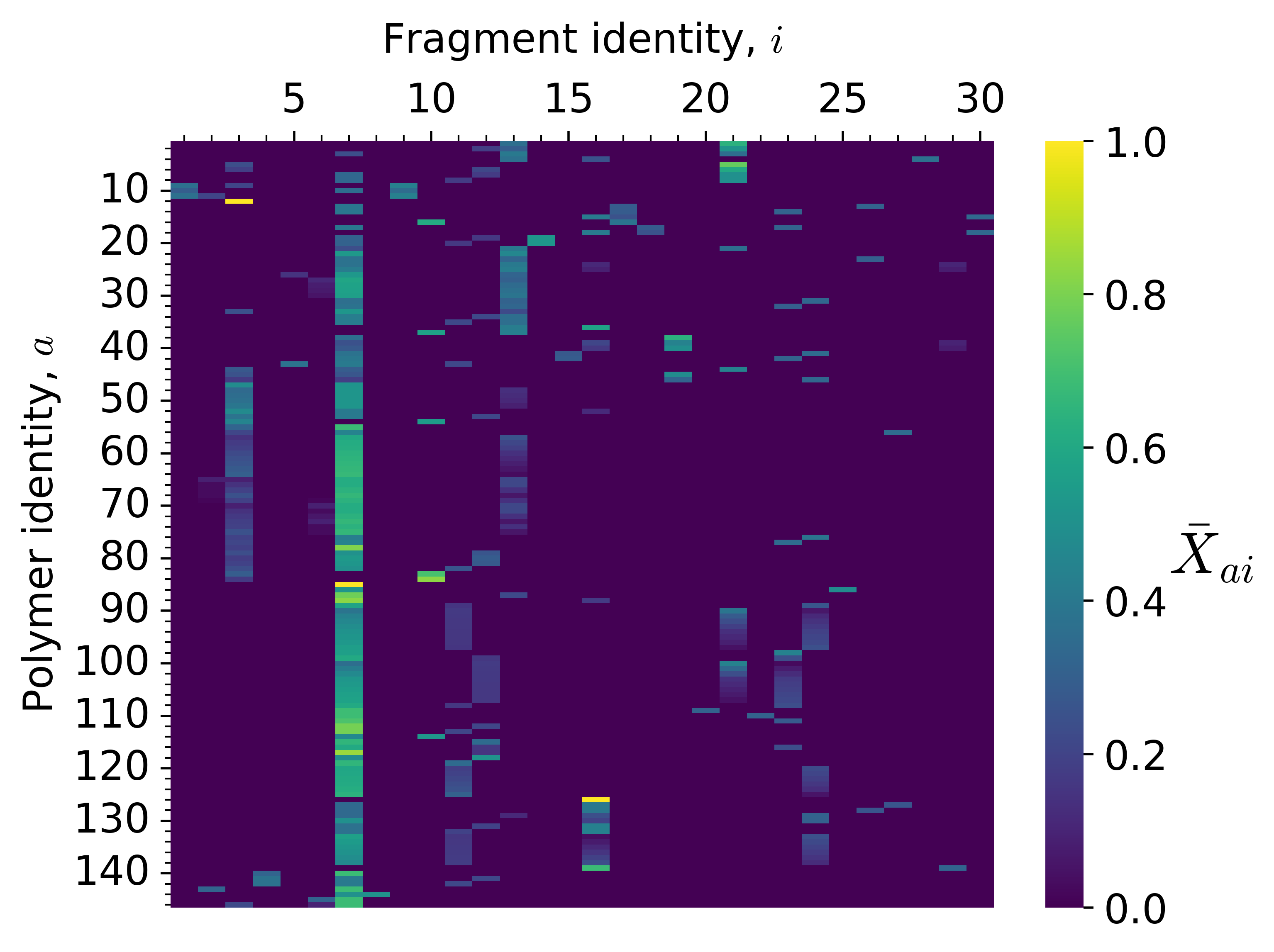}
        \caption{$\X$}
    \end{subfigure}
    \caption{\textbf{Fragment count and composition matrices.} Heat maps of \textbf{(a)} the fragment count matrix $\tensor{X}$ and \textbf{(b)} the molar mass-weighted fragment composition matrix $\X$ calculated from $\tensor{X}$. Plots indicate the degree of sparsity in the matrices, as well as the number and content of fragments per polymer identity. The fragment identity $i$ corresponds to Fig.~\ref{SI_fig:all_fragments_SI}.}
    \label{SI_fig:composition_heatmap}
\end{figure}

\subsection{Model inputs}
 In the Group Additive Properties (GAP) approach the count matrix $\tensor{X}$ is molar mass normalised giving the composition matrix $\X$ with elements $\bar{X}_{ai}$. $\beta_i$ is then estimated from the experimentally available $\Tgval$ values by Ordinary Least Squares (OLS) regression against $\X$.
The column space of $\tensor{X}$, and thus the elements $X_{ai}$, vary depending on the initial definition of a `fragment'. For the GAP analysis, three different versions of $\tensor{X}$ were investigated, corresponding to the fragment definitions: $L\-Ar$, $L\-Ar\-L$, and $Ar\-L\-Ar$, see Fig.~\ref{SI_fig:LRL_motivation}.
Since each polymer is represented as a small subset of the full set of available fragments, the matrix $\tensor{X}$ is sparse. To illustrate this, we present $\tensor{X}$ and $\X$ as a heat map in Fig.~\ref{SI_fig:composition_heatmap} (using the $L\-Ar\-L$ definition); it is clear from the heat map that the ether-phenyl-ketone (E-P-K) fragment ($i=7$) contributes significantly to the polymer composition throughout the data set. In this work, we have focused on only fragment definitions that join the $L$ and $Ar$ groups in the fragment identities. Defining count matrices with $L$ and $Ar$ groups represented separately in the columns of $\tensor{X}$ should be approached with caution. This is because the sequence of alternating $Ar$ and $L$ groups, in the general PAEK monomer structure, leads to the linear constraint that the number~of~$Ar$~groups is equal to the number~of~$L$~groups for every polymer in our data set. 
This linear dependence leads to a non-invertible matrix $\X^\top\X$, which causes issues for the GAP estimation in Eq.~(11) in the main text.

The QSPR-GAP method combines the assumptions of GAP, with quantitative structural properties (QSPR) calculations, requiring the matrix $\X$ to quantify the fragment composition in a given monomer and a descriptor matrix $\D$ to encode the physical properties of the constituent fragments. The information contained in the descriptor matrix $\D$ can be presented in terms of the Pearson correlation matrices $\mathcal{P}^{(\tensor{D},\tensor{R})}_{\mu\nu}$ for $\D$ and the matrix product $\tensor{R}\equiv\X \D$. The Pearson correlation matrix quantifies how different descriptors $\mu$ and $\nu$ are correlated with each other either across the fragments in the data set ($\mathcal{P}^{(\tensor{D})}_{\mu \nu}$) or across  the set of polymers ($\mathcal{P}^{(\tensor{R})}_{\mu \nu}$). Hence,

\begin{equation}
\mathcal{P}^{(\tensor{D})}_{\mu \nu} = \frac{\sum_{i=1}^p (D_{i\mu} - \overline{D}_{\mu}) (D_{i\nu} - \overline{D}_{\nu})}
    {\sqrt{\sum_{i=1}^p(D_{i\mu} - \overline{D}_{\mu})^2} 
    \sqrt{\sum_{i=1}^p(D_{i\nu} - \overline{D}_{\nu})^2}},
\end{equation}
where $\overline{D}_{\mu} \equiv \frac{1}{p}\sum_{i=1}^p D_{i\mu}$ is the mean of descriptor $\mu$ over all fragments $i=1,\ldots,p$. Similarly, 

\begin{equation}
\mathcal{P}^{(\tensor{R})}_{\mu \nu} = \frac{\sum_{a=1}^n (R_{a\mu} - \overline{R}_{\mu}) (R_{a\nu} - \overline{R}_{\nu})}
    {\sqrt{\sum_{a=1}^n(R_{a\mu} - \overline{R}_{\mu})^2
    } 
    \sqrt{
    \sum_{a=1}^n(R_{a\nu} - \overline{R}_{\nu})^2}},
\end{equation}
where $\overline{R}_{\mu} \equiv \frac{1}{n}\sum_{a=1}^n R_{a\mu}$ is the mean of descriptor $\mu$ over all polymers $a=1,\ldots,n$.

\begin{figure}[tbh]
    \centering
    \begin{subfigure}[b]{0.49\textwidth}
        \centering\includegraphics[width=\textwidth]{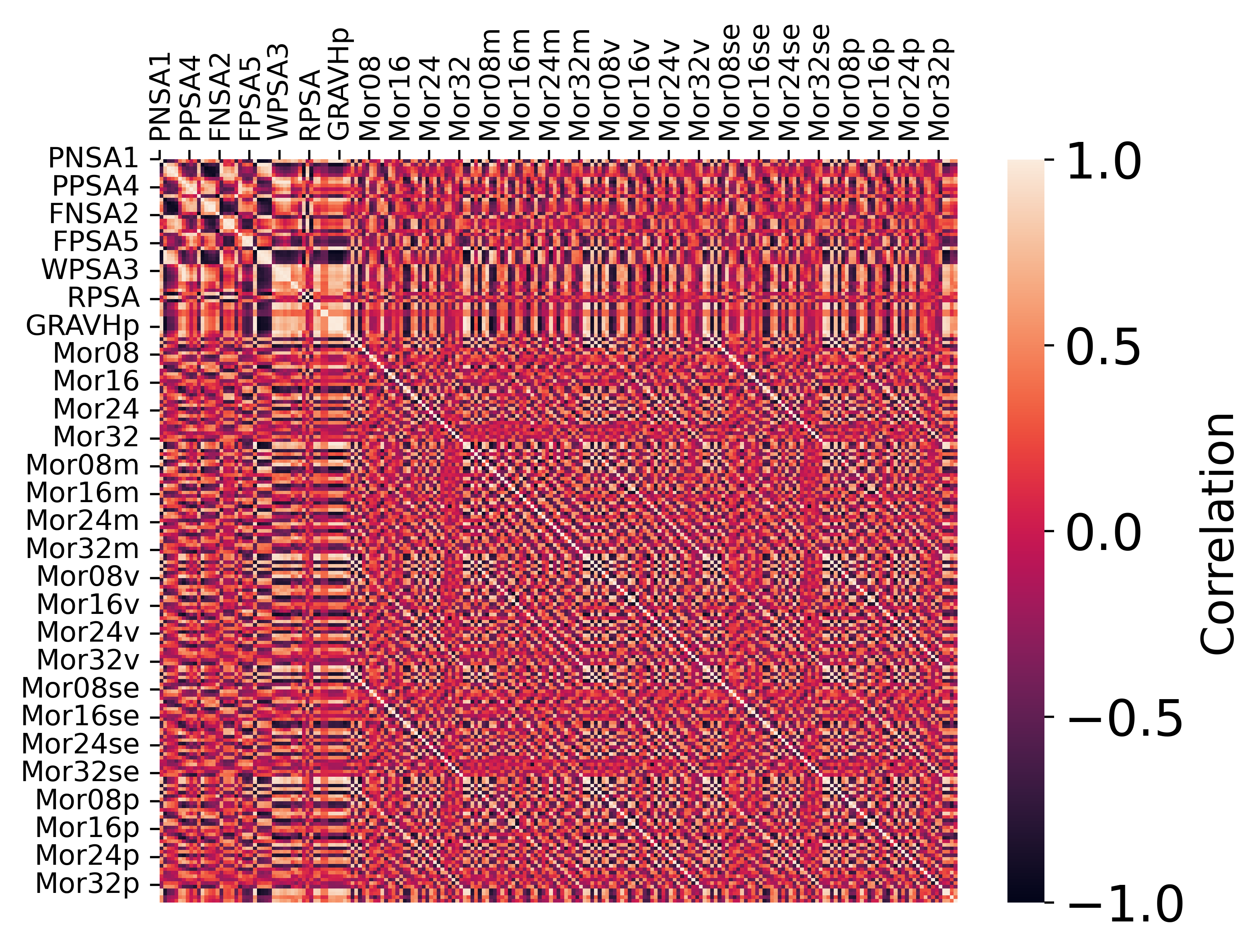}
        \caption{Pearson Correlation matrix of $\mathbf{D}$.}
        \label{subfig:heatmap_D}
    \end{subfigure}
    \hfill
    \begin{subfigure}[b]{0.49\textwidth}
        \centering\includegraphics[width=\textwidth]{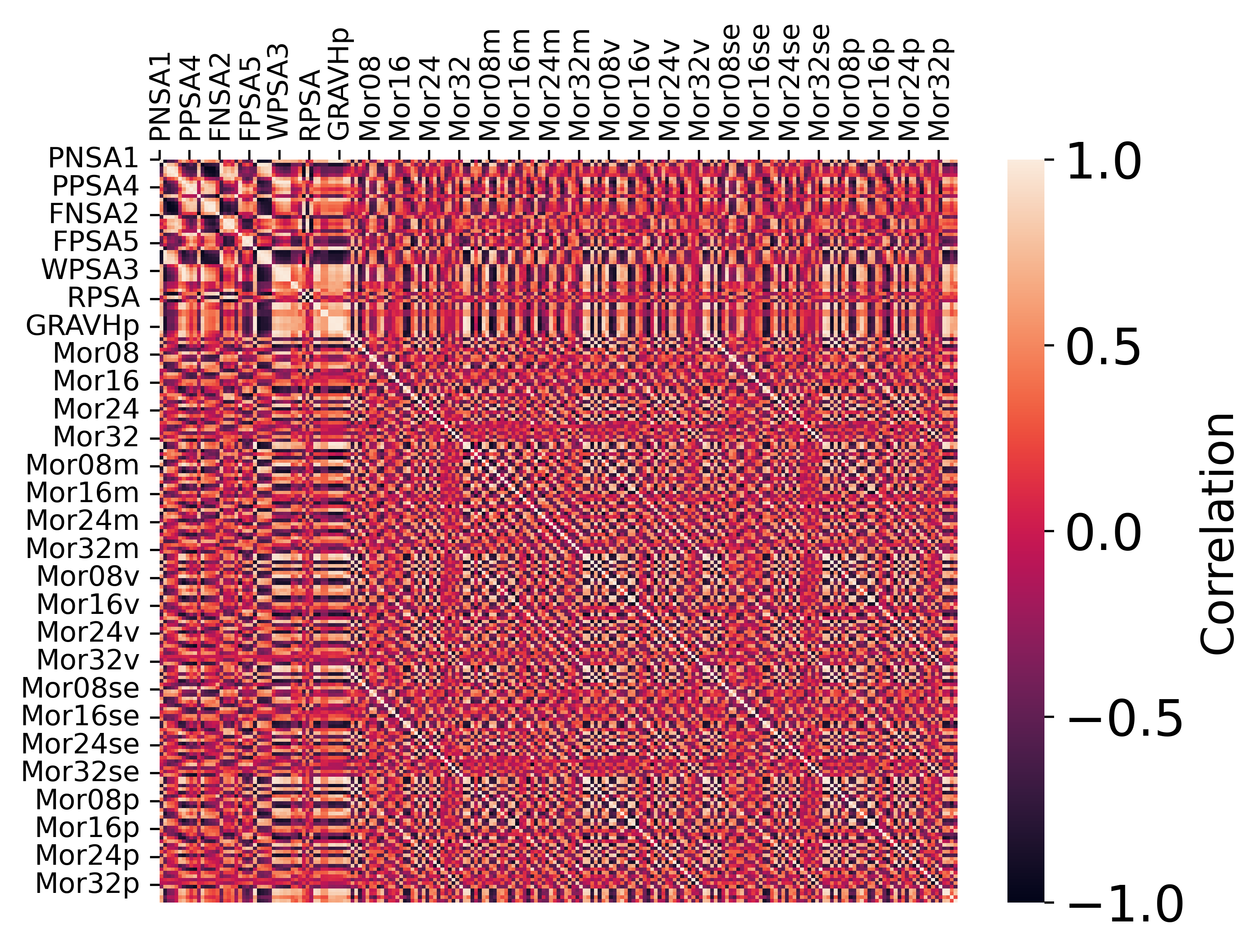}
        \caption{Pearson Correlation matrix of $\X \D$}
        \label{subfig:heatmap_XD}
    \end{subfigure}
    \caption{\textbf{Pearson Correlation Matrices} $\mathcal{P}_{\mu\nu}$ of \textbf{(a)} descriptor matrix $D_{i\mu}$ for all descriptors $\mu=1,\ldots,213$ and summed over the fragments $i=1,\ldots,30$ \textbf{(b)} $R_{a\mu}\equiv (\X \D)_{a\mu}$ for all descriptors $\mu=1,\ldots,213$ and summed over all polymers $a=1,\ldots,146$. 
    The axis labels are displayed at every eighth descriptor, starting from the first, for example, $\mu=1, 9, 17, 25, \ldots$ and so on.}
    \label{SI_fig:descriptor_heatmaps}
\end{figure}

The two Pearson correlation matrices, shown in Fig.~\ref{SI_fig:descriptor_heatmaps},  are very similar. Each correlation matrix has multiple regions of high correlations in the off diagonal elements, and in particular, collections of 3D-MoRSE descriptors (\texttt{Mor01}-\texttt{Mor32p}) exhibit strong correlations for the same $q$ values but different weighting schemes. This is no surprise, since the included weighting scheme (atomic mass, van der Waals volume, electronegativity and polarisability) are strongly correlated on physical grounds. Hence we expect that many fewer descriptors will eventually suffice to capture the behavior of $T_g$.

\begin{figure}[tbh]
    \centering
    \begin{subfigure}[b]{0.45\textwidth}
        \centering\includegraphics[width=\textwidth]{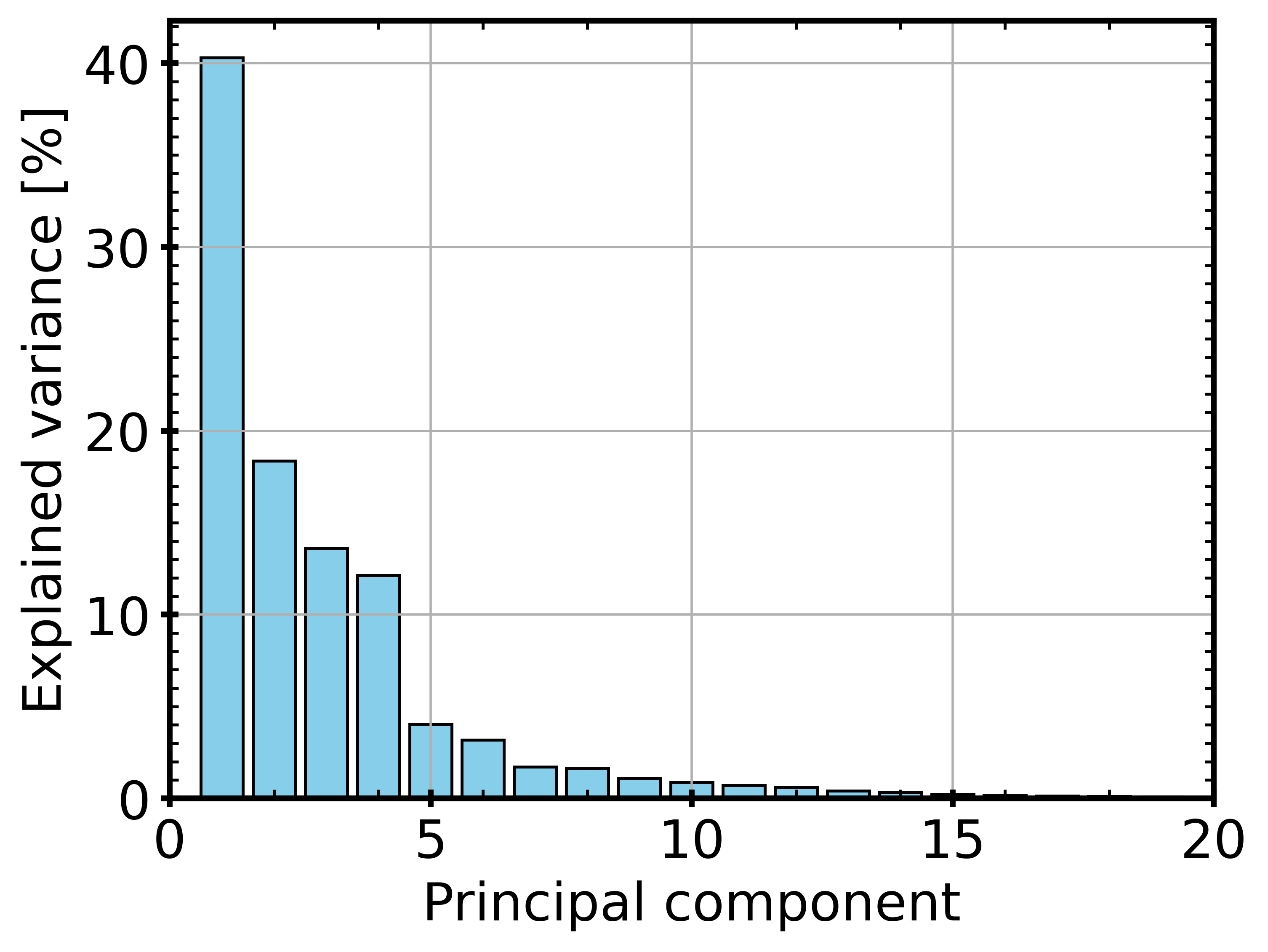}
        \caption{Scree plot of $\mathbf{D}$}
        \label{subfig:scree_D}
    \end{subfigure}
    \hfill
    \begin{subfigure}[b]{0.45\textwidth}
        \centering
        \includegraphics[width=\textwidth]{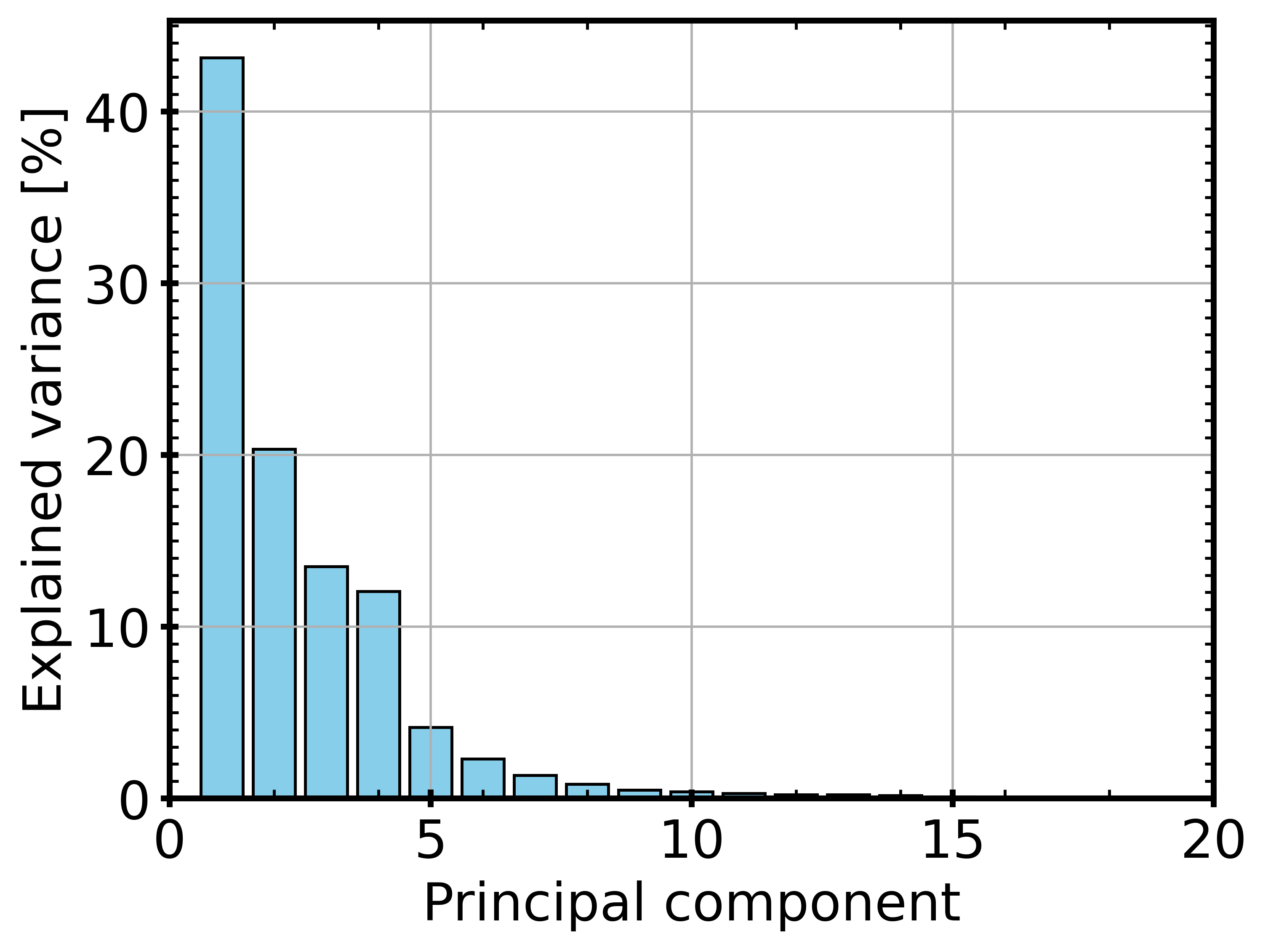}
        \caption{Scree plot of $\X \D$}
        \label{subfig:scree_XD}
    \end{subfigure}
    \caption{\textbf{Principal component analysis of $\D$ and $\R\equiv\X \D$.} The \textit{scree plots} show the eigenvalues of $\D^\text{\!*} \,^{\top}\,\D^\text{\!*}/(p-1)$ and $\R^\text{\!*} \,^{\top} \, \R^\text{\!*}/(n-1)$. The \textit{explained variance} is the ratio between an eigenvalue and the sum of all eigenvalues.}
    \label{SI_fig:scree_plots}
\end{figure}

A principal component analysis (PCA) was performed on the two matrices $\D$ and $\tensor{R}\equiv\X \D$. First, the matrices are standardised to obtain $\D^\text{\!*}$ and $\R^\text{\!*}$ (see Sec.~\ref{SI_Section:standardising})
and then the eigenvalues of $\D^\text{\!*} \,^{\top}\,\D^\text{\!*}/(p-1)$ and $\R^\text{\!*} \,^{\top} \, \R^\text{\!*}/(n-1)$ are computed. Each eigenvalue of $\D^\text{\!*} \,^{\top}\,\D^\text{\!*}/(p-1)$ constitutes the variance of the descriptors corresponding to a particular principle component, over the complete set of fragments. Hence, the total variance within the set of fragments is given by the sum of the eigenvalues. The eigenvalues of $\R^\text{\!*} \,^{\top} \, \R^\text{\!*}/(n-1)$ similarly represent the variance of the mass-weighted descriptors within respective principal components, across the entire set of polymers.  The relative variance captured by each eigenvalue is shown in the \textit{scree plots} (Fig.~\ref{SI_fig:scree_plots}), in which the eigenvalues, normalised by the sum of all eigenvalues, are plotted against the principal component (or eigenvalue) number in order of decreasing magnitude. This is referred to as \textit{explained variance}. The figures show that more than $99\%$ of the variance is captured by the first 15  principal components, despite there being $m=213$ descriptors.

\section{Models}

For the QSPR-GAP method, the ordinary least squares (OLS) method is not appropriate for the estimation of the coefficients $\gamma_0$ and $\gamma_\mu$ ($\mu=1,\ldots,m)$ in the fitting function
\begin{equation}
    f((\tensor{\bar{X}})_a, \tensor{D}) = \gamma_0 + \sum_{i=1}^p \sum_{\mu=1}^m \bar{X}_{ai} D_{i\mu} \gamma_\mu,\label{SI_eq:fitting_func_qspr_gap}
\end{equation}
since this would entail attempting to estimate more coefficients than there are data points ($m=213$ descriptors and the data set contains $n=146$ polymers). In addition, many descriptors in $\D$ are strongly correlated (Fig.~\ref{SI_fig:descriptor_heatmaps}), resulting in multicolinearity between predictors in the design matrix $\tensor{\bar{X} D}$. As demonstrated by the scree plots in Fig.~\ref{SI_fig:scree_plots}, the fact that less than 15 principal components have non-negligible variance for a data set containing a total of 213 descriptors, explains the significant multicollinearity. We therefore employ alternative regression methods, which each introduce a small amount of bias in order to improve the prediction accuracy and address this multicolinearity problem. The used methods are: Ridge Regression, Lasso Regression, Principal Component Regression (PCR) and Partial Least Squares (PLS) Regression \cite{Hastie2001TheLearning}.
\subsection{Standardising variables} \label{SI_Section:standardising}

The regression methods used in this study (PCA, PLS, Ridge and Lasso) are not scale-invariant and solutions can thus differ depending on the units of the inputs. To avoid this, we standardise the inputs to have zero mean and a sample variance of one. The standardised input matrix $\mathbf{R}^{\text{\!*}}$ is calculated as

\begin{align}
R^{\text{*}}_{a\mu} &= \frac{R_{a\mu} - \overline{R_\mu}}{s_{R_\mu}}, && \text{where} & R_{a\mu} &= \sum_{i=1}^p \bar{X}_{ai} D_{i\mu} \equiv (\mathbf{X} \mathbf{D})_{a\mu},
\end{align}
with sample mean $\overline{R_\mu}$ and variance $s_{R_\mu}^2$ for descriptor $\mu$ given by
\begin{align}
\overline{R_\mu} &= \frac{1}{n}\sum_{a=1}^n R_{a\mu}, & s_{R_\mu}^2 &= \frac{1}{n-1}\sum_{a=1}^n (R_{a\mu} - \overline{R_\mu})^2.
\end{align}
Similarly, $\Tga$ is standardised,
\begin{equation}
\Tgval^{\text{*}a} = \frac{\Tga - \overline{\Tgval}}{s_{\Tgval}},
\end{equation}
with sample mean and variance given by
\begin{equation}
\overline{\Tgval} = \frac{1}{n}\sum_{a=1}^n \Tga, \quad s_{\Tgval}^2 = \frac{1}{n-1}\sum_{a=1}^n (\Tga - \overline{\Tgval})^2.
\end{equation}
The standardised fitting function is redefined as
\begin{equation}
f((\mathbf{R}^{\text{\!*}})_{a}) = \sum_{\mu=1}^m R^{\text{*}}_{\,a\mu} \gamma^{\!\text{*}}_{\,\mu},
\end{equation}
where $\gamma^{\!\text{*}}_{\,\mu}$ are the standardised regression coefficients for $\mu=1,\ldots,m$. Once estimated (see below), the regression coefficients $\hat{\gamma}^{\!\text{*}}_{\,\mu}$ can be converted back to the original unstandardised regression coefficient estimations $\hat{\gamma}_\mu$ and $\hat{\gamma}_0$ by
\begin{equation}
\hat{\gamma}_\mu = \frac{s_{\Tgval}}{s_{R_\mu}} \hat{\gamma}^{\!\text{*}}_{\,\mu}, \quad \hat{\gamma}_0 = \overline{\Tgval} - \sum_{\mu=1}^m \hat{\gamma}_\mu \overline{R_\mu}.
\end{equation}

\subsection{Shrinkage methods}

In Ordinary Least Squares the standardised glass transition temperature $\Tgval^{\text{*}a}$ is taken to be linearly related to the standardised matrix $\R^\text{\!*}$, and a least squares minimisation is performed to estimate the standardised regression coefficients $\hat{\g}^\text{\!*}$ according to 
\begin{align}
\hat{\g}^{\!*}_\textit{OLS} &= \underset{\d}{\arg\min} \left\{ \sum_{a=1}^n [\Tgval^{\text{*}a} - \sum_{\mu=1}^m R^{\text{*}}_{\,a\mu} \hat{\gamma}^{\!\text{*}}_{\,\mu}]^2\right\}.   \label{eq_SI:OLS}
\end{align}
For Ridge and Lasso regression methods an additional term (a penalty) with a value related to the size of $\g^{\!\text{*}}$ is incorporated in the function to be minimised. The presence of this term  penalises the size of the estimated coefficients, resulting in many fewer `effective' regression coefficients; this process is known as shrinkage. In Ridge regression the added penalty is proportional to the L2-norm of the coefficients, 
whereas for Lasso regression, the penalty is proportional to the L1-norm of the coefficients, as shown below:

\begin{subequations}
    \begin{align}
    \hat{\g}^\text{\!*}_\textit{Ridge} &= \underset{\g^\text{\!*}}{\arg\min} \left\{ \sum_{a=1}^n [\Tgval^{\text{*}a} - \sum_{\mu=1}^m R^{\text{*}}_{\,a\mu} \gamma^{\text{\!*}}_{\,\mu}]^2 +
    \alpha \sum_{\mu=1}^m (\gamma^{\text{\!*}}_{\,\mu})^2
    \right\}, \label{eq_SI:ridge}\\
    \hat{\g}^\text{\!*}_\textit{Lasso} &= \underset{\g^\text{\!*}}{\arg\min} \left\{ \sum_{a=1}^n [\Tgval^{\text{*}a} - \sum_{\mu=1}^m R^{\text{*}}_{\,a\mu} \gamma^{\text{\!*}}_{\,\mu}]^2 +
    \alpha \sum_{\mu=1}^m \lvert \gamma^{\text{\!*}}_{\,\mu} \rvert
    \right\}, \label{eq_SI:lasso}
    \end{align}
\end{subequations}
where the hyperparameter $\alpha$ controls the degree of applied coefficient shrinkage. If $\alpha$ is sufficiently large, then for Lasso regression (L1) some coefficients will shrink to exactly zero, whereas for Ridge regression, some coefficients will shrink to values close to zero but never actually reach zero; this is due to the nature of each penalty term (for further information, see \cite{Hastie2001TheLearning}). As $\alpha$ approaches zero, both Ridge and Lasso regression will converge to the OLS solution. The hyperparameter $\alpha$ is optimised during internal validation, which is discussed further in Section~\ref{SI_sec:model_evaluation}.

\subsection{PCR and PLS dimension reduction methods} \label{SI_Section:Dim_reduction}

Both Principal Component Regression (PCR) and Partial Least Squares (PLS) regression consist of a transformation of the standardised input data $\R^{\text{\!*}}$, and the application of a regression procedure. PCR executes these steps separately, while PLS performs them simultaneously. Both regression methods use the \textit{singular value decomposition} (SVD) factorisation on the non-square $n\times m$ matrix $\R^{\text{\!*}}$. If $\R^{\text{\!*}}$ has rank $r$ then it has $r$ non-zero singular values, which equivalently correspond to the $r$ non-zero (square roots of) eigenvalues of the matrix $\R^{\text{\!*}}\,^\top\,\R^{\text{\!*}}$.
The SVD of $\R^{\text{\!*}}$ is given by $\R^{\text{\!*}} = \U \, \S \, \V^\top$ where $\U_{n\times r}$ and $\V_{m \times r}$ are respectively the sets of left and right singular vectors of $\R^{\text{\!*}}$, and $\S_{r \times r}$ is the diagonal matrix of singular values, ordered from largest to smallest.

PCR transforms and compresses $\R^{\text{\!*}}$ onto its first $k$ principal components ($k \leq r$), using only the first $1, \ldots, k$ columns of $\V$, denoted $\V_k$ ($m \times k$).
The  transformation $\R^{\text{\!*}} \, \V_k$ drops the $(k+1), \ldots, m$ columns of $\R^{\text{\!*}} \, \V$ that explain the smallest sample variance (corresponding to the smallest eigenvalues of $\R^{\text{\!*}}\,^\top\,\R^{\text{\!*}}$). Then, a least squares minimisation is performed on the transformed matrix $\R^{\text{\!*}} \, \V_k$ with $k$ regression coefficients $\boldsymbol{\xi} \in \mathbb{R}^k$, yielding the estimator
\begin{equation}
\hat{\boldsymbol{\xi}} = \underset{\boldsymbol{\xi}}{\arg\min} \lVert \Tgvec^\text{\!*} - \R^\text{\!*} \, \V_k \, \boldsymbol{\xi} \rVert^2.
\label{SI_eq:PCR_minimisation}
\end{equation}
The first $k$ components are referred to as $n_\text{components}$ in Table~\ref{SI_table:hyperparameter_ranges}.
The solution to Eq.~\eqref{SI_eq:PCR_minimisation} gives
\begin{equation}
\hat{\boldsymbol{\xi}}  = \S_k^{-1} \, \U_k ^{\,\top} \, \Tgvec^\text{\!*},
\end{equation}
since $\R^\text{\!*} \, \V_k = \U_k \, \S_k$, where $\S_k$ is the diagonal matrix of the largest $k$ singular values ($k\times k$) and $\U_k$ is the set of left singular vectors ($n \times k$). Importantly, the reduced standardised coefficient estimates $\hat{\g}^\text{\!*}_\textit{PCR}$ are determined by the transformation $\hat{\g}^\text{\!*}_\textit{PCR}=\V_k\, \hat{\boldsymbol{\xi}}$ \cite{Jolliffe2002PrincipalAnalysis}.
PCR will perform poorly if $\Tgvec^\text{\!*}$ is strongly correlated in directions with small variance, \textit{i.e.} principal components that are strongly related to $\Tgval$, but have small singular values. 

PLS does not suffer this issue since it includes $\Tgvec^\text{\!*}$ in the data compression process. Like PCR, the PLS algorithm compresses the data to the first $k$ components 
($n_\text{components}$ in Table~\ref{SI_table:hyperparameter_ranges}), but now using the first $k$ left and right singular vectors from the SVD of the inner product of $\R^{\text{\!*}} {^\top} \, \Tgvec^\text{\!*}$. We use the formulation of PLS from the Python module Scikit-learn \cite{scikit-learn}, which is implemented from \citet{PLS2000}. 

\subsection{Genetic algorithm}

A genetic algorithm employs concepts inspired by biology to solve optimisation problems, utilising processes such as natural selection, random mutation, and genetic recombination \cite{Sukumar2014}.
We follow the procedure outlined in Fig.~\ref{SI_fig:GA_flow_diagram} \cite{Gasteiger2003HandbookChemoinformatics}. 
In the present study, a chromosome is represented as a binary bit string (Fig.~\ref{SI_fig:chromosome}) of $m=213$ genes, where the $\mu$th gene corresponds to the $\mu$th descriptor. The value 1 signifies that a gene (descriptor) is `turned on', while the value 0 means that a gene (descriptor) is not turned on.

\begin{minipage}{0.4\textwidth}
\begin{figure}[H]
\centering\includegraphics[width=0.8\textwidth]{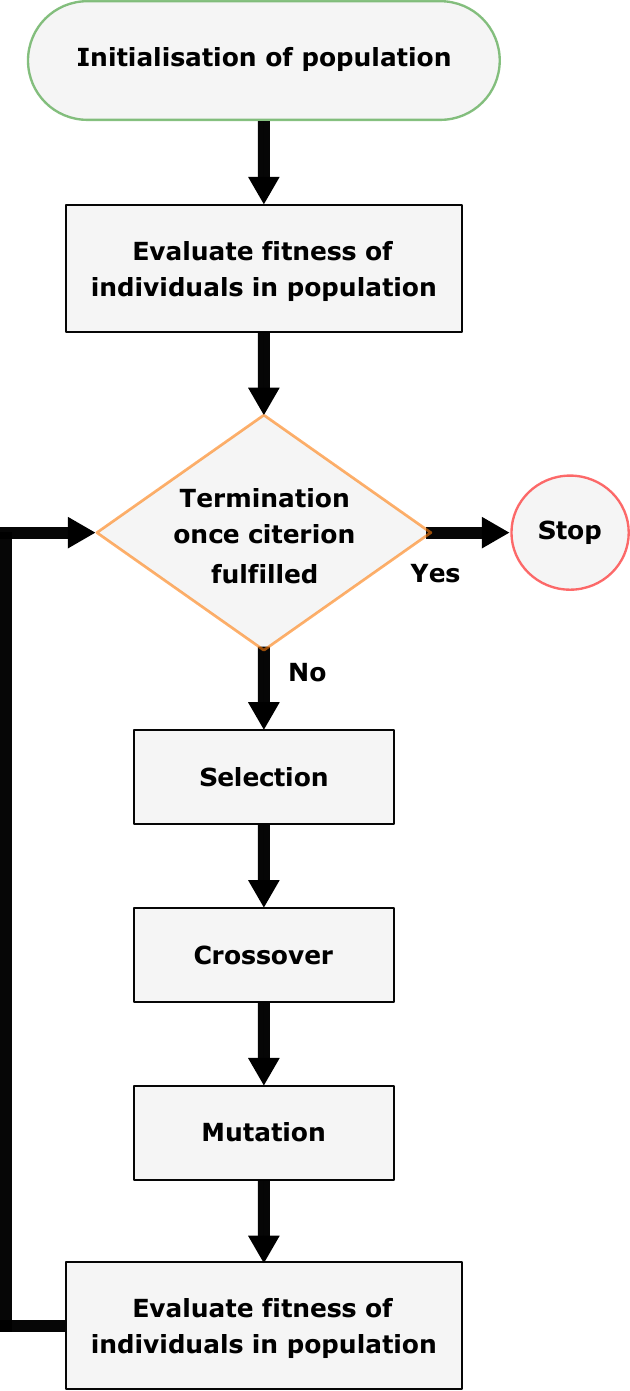}
    \caption{\textbf{Genetic algorithm flow diagram,} starting with an initialisation of a population and ending with a termination criterion. Each step is discussed in the current section and the parameters accompanying each step can be found in Table~\ref{SI_table:GA_parameters}.}\label{SI_fig:GA_flow_diagram}
\end{figure}
\end{minipage}\hfill
\begin{minipage}{0.55\textwidth}
\begin{table}[H]
    \centering
    \rowcolors{2}{white}{gray!25} 
    \begin{tabular}{>{\raggedright\arraybackslash}p{4cm}p{5.2cm}}
        \hline\hline
        \textbf{GA parameter} & \textbf{Value} \\
        \hline\hline
        Population size & 213 chromosomes \\
         Gene pool & 213 genes (molecular descriptors) \\
        Number of generations & 50 generations \\
        \texttt{mutation\_rate\_per\_pop} & $\text{probability}=0.1$ \\
        \texttt{mutation\_rate\_per\_chrom} & $\text{proportion}=0.9/\mG$ \\
        \hline\hline
\end{tabular}
\caption{\textbf{Parameter inputs for the genetic algorithm.} Population size is measured in the number of chromosomes, which is taken to match the size of the gene pool, and the algorithm terminates after the number of generations. The mutation rate per population, given by \texttt{mutation\_rate\_per\_pop}, defines the probability a chromosome will be mutated in the population (in a given generation). 
The mutation rate per chromosome is defined by a randomly selected subset of genes that are randomly shuffled. The value of \texttt{mutation\_rate\_per\_chromosome} defines the ratio of this subset size to the size of the chromosome.}\label{SI_table:GA_parameters}
\end{table}
\begin{figure}[H]
    \centering\includegraphics[width=\textwidth]{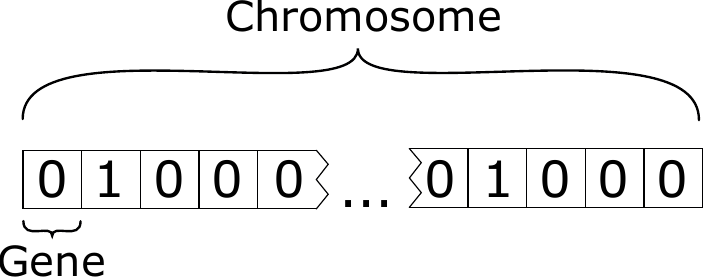}
    \caption{\textbf{Binary representation of a chromosome.} Chromosomes used in the present study contain 213 genes, where the $\mu$th gene takes values 0 (turned off) or 1 (turned on) and corresponds to the $\mu$th descriptor/feature $(\D)_\mu$.}\label{SI_fig:chromosome}
\end{figure}
\end{minipage}
\vskip0.3truecm

The algorithm begins with a set of initial conditions, as shown in Table~\ref{SI_table:GA_parameters}, the first being \textit{population size}.  A population consists of a number of chromosomes, where each chromosome is represented by a sequence of randomly distributed 1's and 0's (genes).  The population size is set to be equal to the size of the gene pool, \textit{i.e.} the number of genes in a single chromosome. This enables enough variation of genes to represent the full set of descriptors when the population is initialised. Despite the random distribution of 1's and 0's throughout a chromosome, all chromosomes are constrained to have a fixed number of $\mG$ active genes (value 1). For example, QSPR-GAP GA2 always contains 2 genes with value 1, and the remaining 211 genes have value 0.

\begin{figure}[htb]
    \centering\includegraphics[width=.55\textwidth]{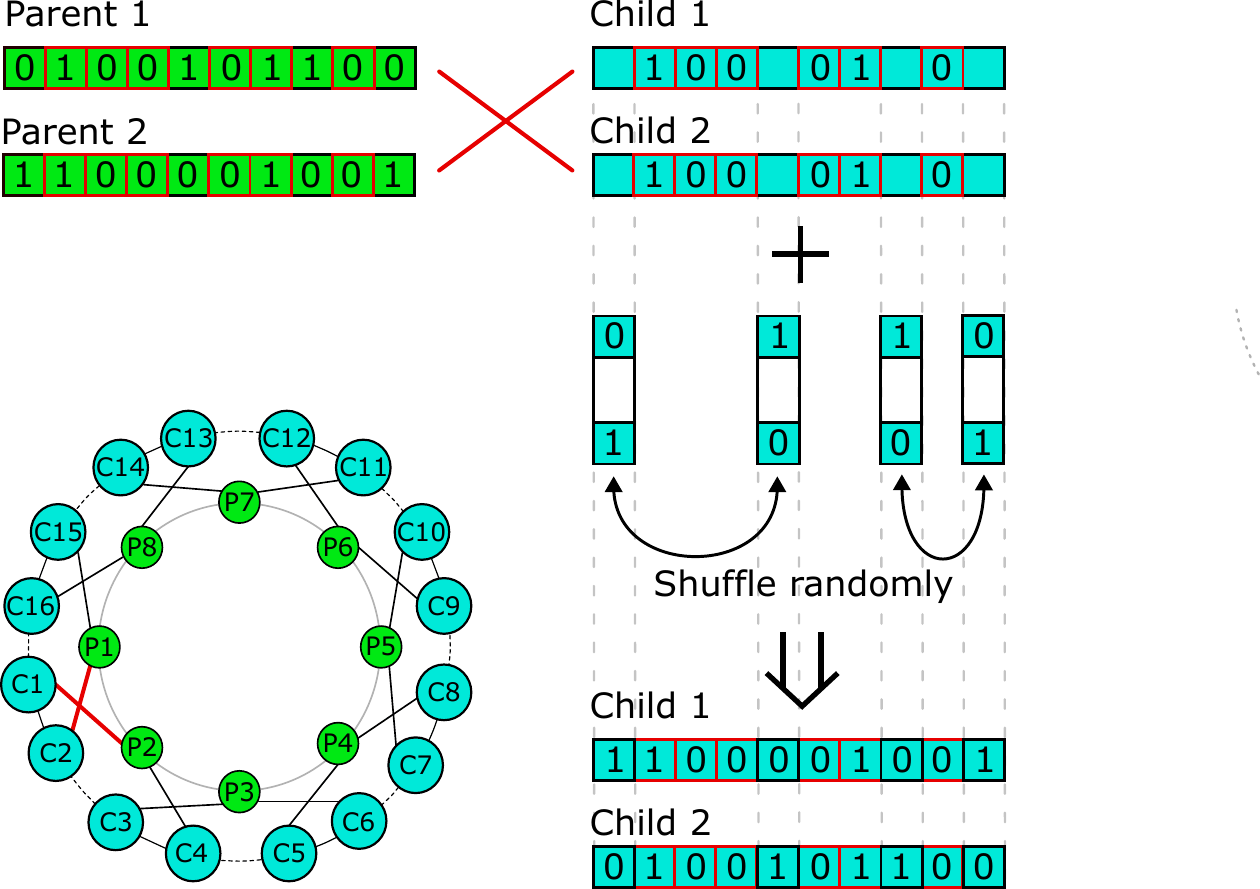}
    \caption{\textbf{The crossover operation}. In this process each parent chromosome combines with \textit{two} other parents to produce a total of  two child chromosomes per parent. The top and the right show P1 and P2 sharing genes to children C1 and C2. The crossover wheel  (bottom left) indicates how the crossover operation is applied to the population as a whole, so that P1 crosses over with P8 (creating C15 and C16) as well as P2, while P2 crosses over with P3 (creating C3 and C4) as well as with P1. This doubles the population.}
    \label{SI_fig:crossover_mechanism}
\end{figure}

The initial population is evaluated for each individual chromosome's \textit{fitness}. The fitness is evaluated using a fitness function which is determined during internal validation (see Fig.~\ref{SI_fig:external_internal_CV_GA}). The genes are indexed according to the molecular descriptors $\mu$. When the $\mu$th gene = 1, then the $\mu$th descriptor is incorporated into a linear regression model. When more than one gene is turned on (equal to 1), then the corresponding descriptors are used in a multiple linear regression; specifically, the robust regression model \texttt{sklearn.linear\_model.HuberRegressor} was used (with Scikit-learn's default hyperparameters). To determine the fitness, consider a single training-test split for a single chromosome within the internal validation. Through some random process a set of $\mG$ descriptors have been `activated' within the chromosome. These descriptors are then included in the regression model, where the training set is used to fit the model and predictions are made on the test set (still within the internal validation). Once the prediction has been made the root-mean-squared-error (RMSE) is calculated between the predictions and test data (see Eq.~\ref{SI_eq:RMSE}). The fitness of the $i$th chromosome was defined as $f_i = 1/\text{(RMSE)}_i$ and every chromosome is assigned a fitness values accordingly.

Based on the fitness of the chromosomes, 50\% of the population is then \textit{selected} to have their genes passed on to the next generation. The mechanism used was \textbf{roulette wheel selection} \cite{Gasteiger2003HandbookChemoinformatics,Sukumar2014} (RWS) which introduces a stochastic element to the selection process. Given a population of $N$ chromosomes, the probability of selecting the $i$th chromosome is distributed according to $p_i=f_i/\sum_{i=1}^Nf_i$. The next generation is created by sampling $N/2$ chromosomes according to this distribution with replacement; \textit{i.e.} the same chromosome can be sampled more than once for the next generation.

The genetic recombination step, known as \textit{crossover}, consists of two parent chromosomes mixing their genes into two child chromosomes, but with each parent chromosome undergoing crossover with two different parent chromosomes to yield two children for every parent, on average. In the example shown in Fig.~\ref{SI_fig:crossover_mechanism}, Parents 1 and 2 (P1 and P2) crossover to make children 1 and 2 (C1 and C2). However P1 also crosses over with P8 to make C15 and C16, while P2 crosses over with P3 to make C3 and C4. In this manner the population doubles, which exactly counteracts the halving of the population that occurred during the selection phase. To ensure that the total number of genes equal to 1 (turned on) in a given chromosome ($\mG$) remains constant, the crossover mechanism adheres to the method illustrated in Fig.~\ref{SI_fig:crossover_mechanism}. For illustrative purposes, we only show ten genes per chromosome in Fig.~\ref{SI_fig:crossover_mechanism} (when actually there are 213 genes). The crossover process is governed by the following rules: when the $\mu$th gene is the same for both P1 and P2 (as indicated by the red boxes around the genes) then this `strong' $\mu$th gene passes on to C1 and C2 with probability 1. For the remaining genes which do not match, these are (randomly) shuffled in any order as shown in the figure. This process guarantees the $\mG$ `turned on' genes remains the same in both C1 and C2 as was previously in P1 and P2.

Once the next generation has been created, the resulting child chromosomes are \textit{mutated} to ensure diversity through the population and prevent premature convergence to local optima. The probability of mutation is controlled by two parameters: \texttt{mutation\_rate\_per\_pop} and \texttt{mutation\_rate\_per\_chrom} which both take values between 0 and 1. The former controls the probability that a chromosome in the population will be mutated during a given round of mutation. If $\texttt{mutation\_rate\_per\_pop}=0.5$ then there is a 50\% chance a given chromosome will be mutated. The latter controls the proportion of genes in the chromosome that will be randomly shuffled; \textit{i.e.} if $\texttt{mutation\_rate\_per\_chrom}=0.1$ then 21 randomly chosen genes will be randomly shuffled, out of the total 213 genes in a chromosome. Since the chromosomes are dominated by zeros ($\mG\leq10$), the selection defined by $\texttt{mutation\_rate\_per\_chrom}$ will often just shuffle zeros and not change the chromosome. Hence, the effective mutation rate is roughly proportional to the number of genes ($\mG$) that have been turned on. To counter this, we have assumed that $\texttt{mutation\_rate\_per\_chrom}$  is inversely proportional to $\mG$, to enhance the possibility of mutations when just a few genes are turned on (Table~\ref{SI_table:GA_parameters}); this is  an attempt to maintain a level of independence between the mutation rate per chromosome and the total number of active genes $\mG$.

Finally, the fitness of the new generation is evaluated and the process repeats until \textit{termination}, which we have specified as after 50 generations. At this point, the best solution (\textit{i.e.} chromosome with best fitness) from all 50 generations was used for the model. The set of optimised descriptors corresponding to the  best (fittest) chromosome were used in the \textit{external validation}, in which those optimised descriptors were fit to the training set to finally make predictions about the test data (see Fig.~\ref{SI_fig:external_internal_CV_GA}).

\section{Model evaluation} \label{SI_sec:model_evaluation}
For all models, we perform an \textbf{external validation} and an \textbf{internal validation}.  The goal of the external validation is to predict polymer $\Tgval$'s that were not used to determine/train the model parameters (`out-of-sample data'), while the goal of the internal validation was to tune hyperparameters for a given model, or to select the optimal set of descriptors. For clarity, we refer to the QSPR-GAP PCR, Ridge, Lasso and PLS models as the \textbf{`Statistical models'}; and 
models that use descriptor selection based on a genetic algorithm as \textbf{`Genetic Algorithm Models'.}
The internal validation was conducted differently for these two sets of approaches (see Fig.~\ref{SI_fig:external_internal_CV_stats} and Fig.~\ref{SI_fig:external_internal_CV_GA}), but the external validation was the same for both. For the GAP models, no internal validation was required.

The external validation was performed using a repeated five-fold cross validation (5-fold CV), where the full data set was shuffled randomly and subsequently partitioned into five exclusive subsets. A test set was iteratively selected from the 5 subsets and, in each iteration, the remaining four subsets were concatenated into a training set. The partitioning into five subsets was repeated 10 times. This procedure resulted in a total of 50 different train-test splits with 50 different combinations of polymers in the training and test sets. For every train-test split, the test set was left out and the internal validation was conducted on the remaining training data set, where the models were tuned and optimised.

The performance metric used for validation (both internal and external) was the root-mean-squared error of the test data, given by
\begin{equation}
    \text{RMSE} = \sqrt{\frac{\sum_{b=1}^B (\Tgb - \Tgbhat)^2}{B}},
    \label{SI_eq:RMSE}
\end{equation}
where $b=1,\ldots,B$ indexes over the polymers in the test set. $\Tgb$ is the experimental $\Tgval$ for the $b$th polymer and $\Tgbhat$ is an out-of-sample $\Tgval$ prediction made on the $b$th polymer.

\newpage
\subsection{Statistical models}
\begin{figure}[H]
    \centering
    \includegraphics[width=.85\textwidth]{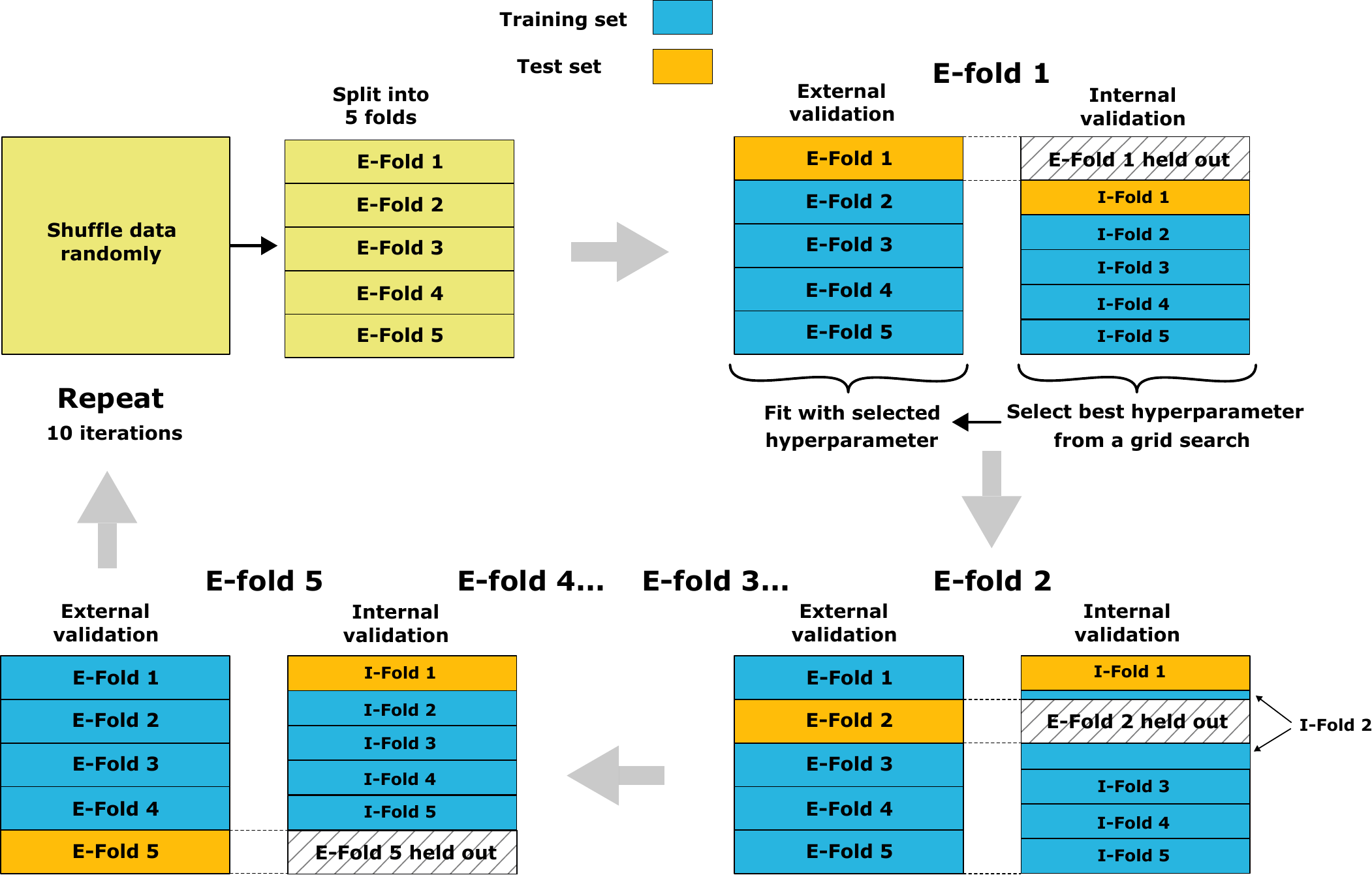}
    \caption{\textbf{Model validation for the `statistical models'.} The key take-away of the figure is the partitions of the data during external and internal validation. When a test set is selected (iteratively) during external validation, this test set is \textbf{held out} from the internal validation as indicated by the diagram. The optimal hyperparameters are selected through internal validation, after which they are used for the external validation. In this phase (external), the model is fit with the training data (and optimised hyperparameters), and used to predict the out-of-sample (test) data; the performance accuracy measure used is the RMSE (Eq.~\ref{SI_eq:RMSE}).
    }
    \label{SI_fig:external_internal_CV_stats}
\end{figure}

\begin{figure}[h]
    \centering
    \includegraphics[width=.8\textwidth]{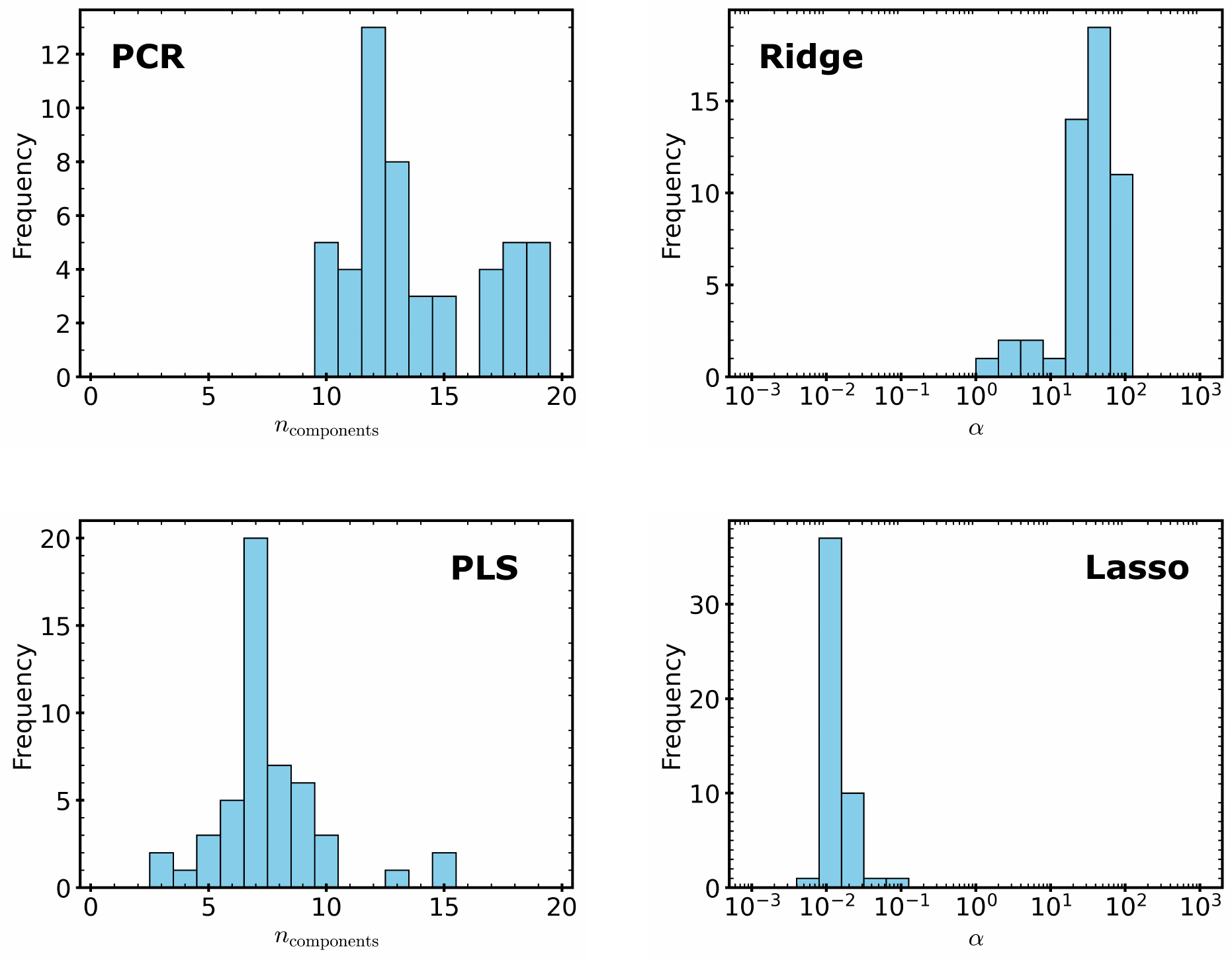}
    \caption{\textbf{Distributions of optimised hyperparameters,} showing the number of occasions a hyperparameter is picked from the range in Table~\ref{SI_table:hyperparameter_ranges}. The hyperparameters are selected based on the optimisation process during internal validation. For a given training-test split (in the external validation), the optimal hyperparameter is selected by applying the internal validation on the external training set. Overall there are 50 splits, and thus 50 different optimal hyperparameters (for the different data partitions). We show the distribution of these 50 hyperparameters here.}
    \label{SI_fig:hyperparameter_distributions}
\end{figure}

For the internal validation of the statistical models, hyperparameter tuning was performed using a one dimensional exhaustive grid search with Scikit-learn's function \texttt{sklearn.model\_selection.GridSearchCV}. The grids of values for each hyperparameter
are listed in Table~\ref{SI_table:hyperparameter_ranges}. During the internal validation, the training data was split into five folds (Fig.~\ref{SI_fig:external_internal_CV_stats}). All discrete values defined in Table~\ref{SI_table:hyperparameter_ranges} were applied (as appropriate for each model) using 4 folds to fit the model (the internal training set) and evaluated on the remaining fold (internal test set). This process was repeated for each of the five folds, and the hyperparameter which yielded the best performance (lowest RMSE) of the five folds was selected for the external validation. The optimised hyperparameters for all 50 training-test splits are presented in Fig.~\ref{SI_fig:hyperparameter_distributions}; these represent the best hyperparameters selected from the 50 internal validations, which were subsequently used for the external validation.

\begin{table}[htb]
    \centering
    \rowcolors{2}{white}{gray!25} 
    \begin{tabular}{>{\raggedright\arraybackslash}cp{3cm}p{4.1cm}p{5.2cm}}
        \hline\hline
        \textbf{Estimator} & \textbf{Hyperparameters} & \hfil\textbf{Range}\hfil &  \hfil\textbf{Python Command} \hfil\\
        \hline\hline
        PCR & $n_{\textrm{components}}$ & Integers $\mathbb{Z}=1,\ldots,20$ & \texttt{list(range(1, 21))}\\
        Ridge & $\alpha$ & $\log_{10}\alpha\in[-3,3]$ in 21 steps & \texttt{list(numpy.logspace(-3, 3, 21))} \\
        Lasso & $\alpha$ & $\log_{10}\alpha\in[-3,3]$ in 21 steps &\texttt{list(numpy.logspace(-3, 3, 21))}\\
        PLS & $n_{\textrm{components}}$ & Integers $\mathbb{Z}=1,\ldots,20$  & \texttt{list(range(1, 21))}\\
        \hline\hline
    \end{tabular}
    \caption{\textbf{Candidate hyperparameter values for grid search implementation.}} 
    \label{SI_table:hyperparameter_ranges}
\end{table}

\newpage

\newpage
\subsection{Genetic algorithm models}

\begin{figure}[H]
    \centering\includegraphics[width=0.85\textwidth]{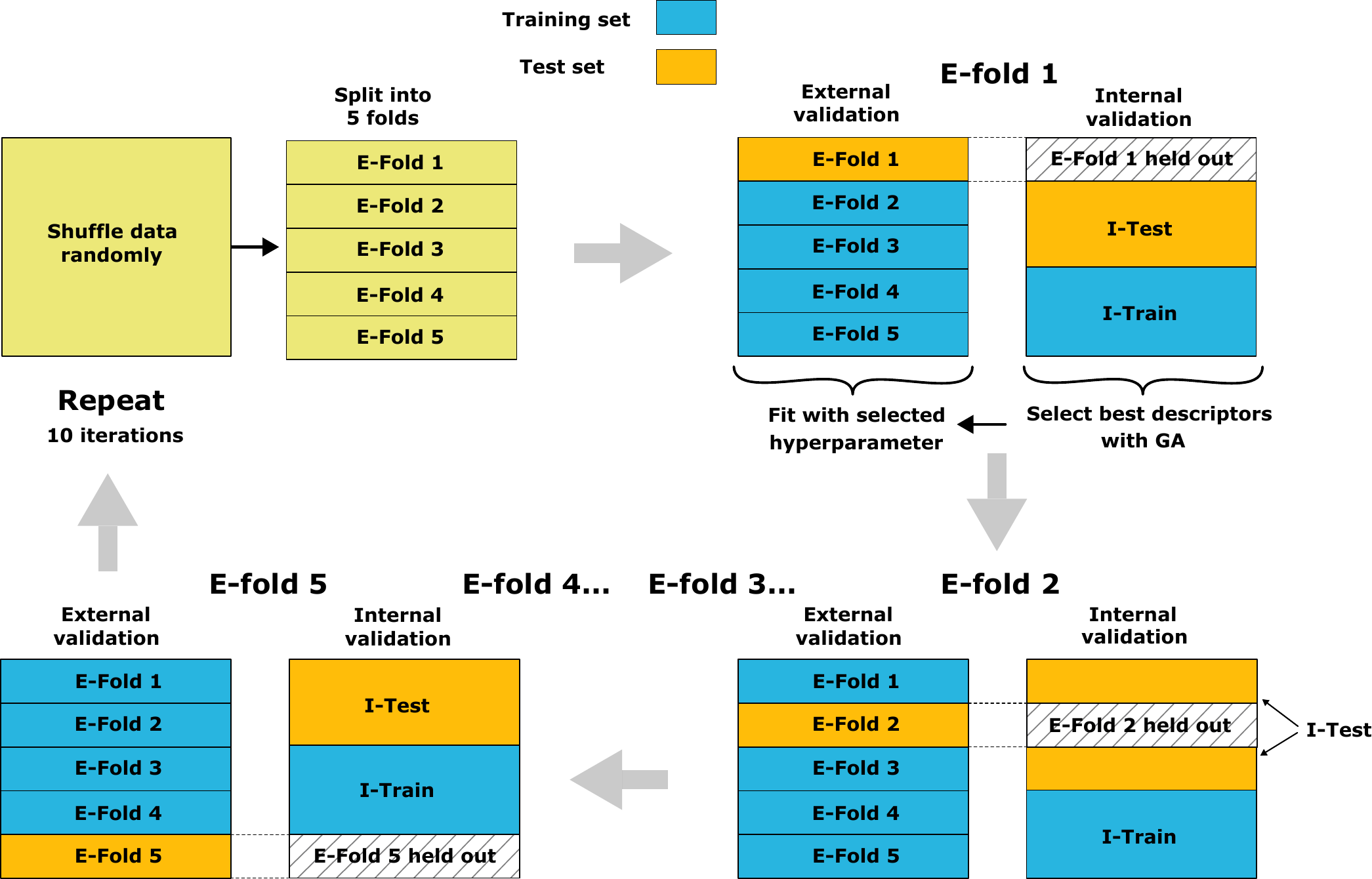}
    \caption{\textbf{Model validation for the `genetic algorithm models'.} The division of training and test sets in both external and internal validations is shown. For a given external training set, a two-way split of internal training and test data was used to calculate the fitness (inverse RMSE). In total there are 50 different two-way splits, each representing a unique internal validation. Within each internal validation, the genetic algorithm evolved over 50 generations of chromosomes to optimise the fitness (see Fig.~\ref{SI_fig:GA_fitness_vs_gen}). The internal validation thus outputs the fittest chromosome (lowest RMSE), corresponding to the optimal $\mG$ descriptors. These descriptors were subsequently used for the external validation, predicting the test data from the training data with OLS; the predictive accuracy was measured with the RMSE (Eq.~\ref{SI_eq:RMSE}).}
    \label{SI_fig:external_internal_CV_GA}
\end{figure}

\begin{figure}[H]
    \centering\includegraphics[width=\textwidth]{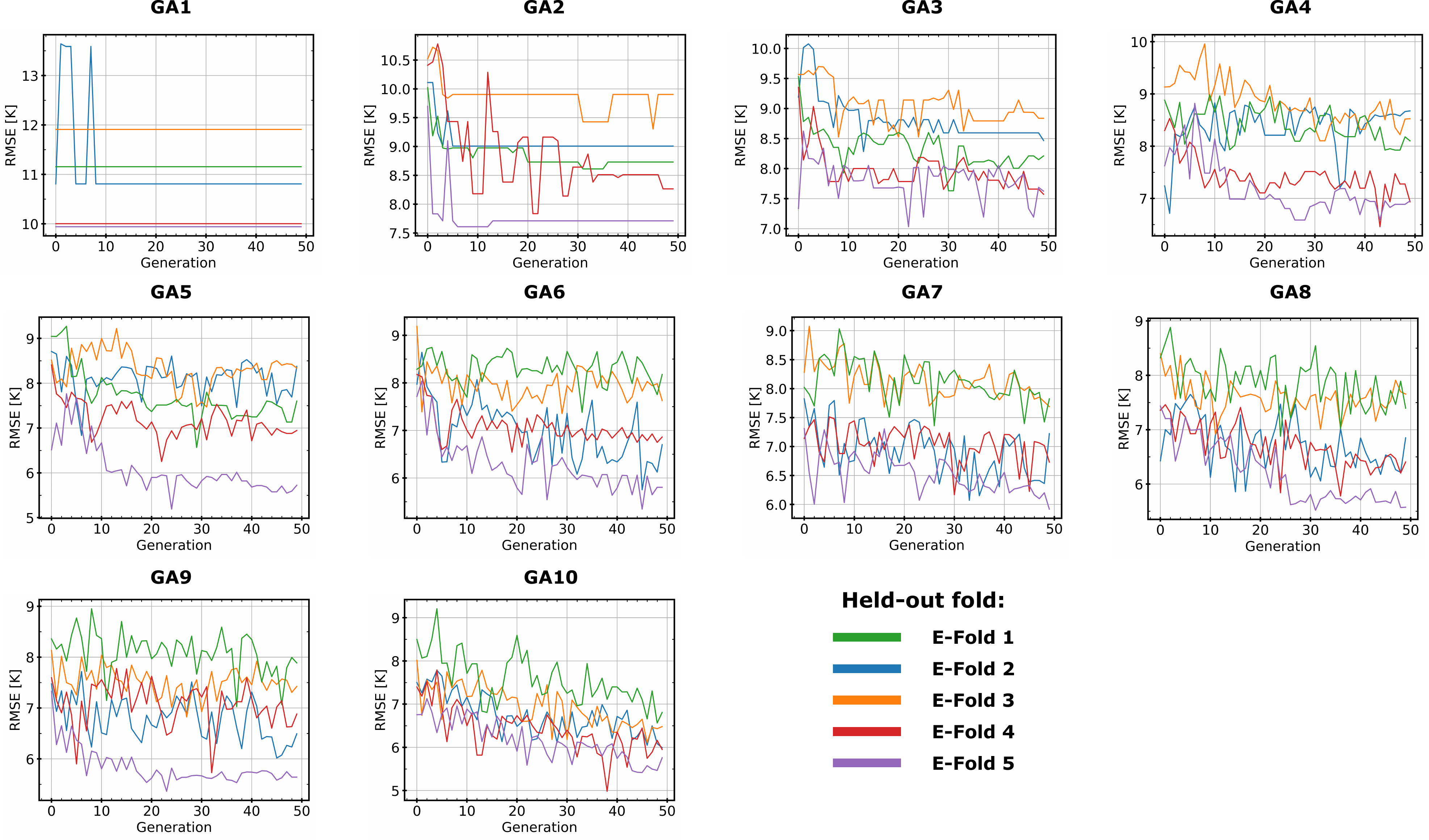}
    \caption{\textbf{Evolution of chromosomes for all genetic algorithm models.} The RMSE of the best chromosome per population is shown through each generation.
    The RMSE is calculated in the internal validation (Fig.~\ref{SI_fig:external_internal_CV_GA}) and the chromosome with the lowest RMSE (from all generations) is then applied in the external validation to predict the held-out fold (shown by the plot legend). All ten GA$\mG$ models are presented, where $\mG$ is the number of features/descriptors used in each model (the number of genes `turned on' to 1 in a chromosome).}
    \label{SI_fig:GA_fitness_vs_gen}
\end{figure}

For the GA models, the internal validation included 50\% training and 50\% test data selected at random (Fig.~\ref{SI_fig:external_internal_CV_GA}). The GA was applied during the internal validation; the fitness of the chromosomes was evaluated by fitting to the internal training set and predicting the internal test set, using the inverse RMSE as the fitness metric. Once the best chromosome was determined by maximising the fitness, the set of descriptors corresponding to the best chromosome were used to fit the model to the training data in the external validation, and to predict the test data. This process was repeated a total of 50 times as depicted in Fig.~\ref{SI_fig:external_internal_CV_GA}.
The evolution of fitness through each generation was recorded for all 50 of the internal validations conducted. We show an example of five from the 50 internal validations in Fig.~\ref{SI_fig:GA_fitness_vs_gen} for the ten different QSPR-GAP GA$\mG$ models ($\mG = 1, \ldots, 10$). 

\subsection{Summary of results}
\begin{figure}[H]
    \centering
    \includegraphics[width=.35\textwidth]{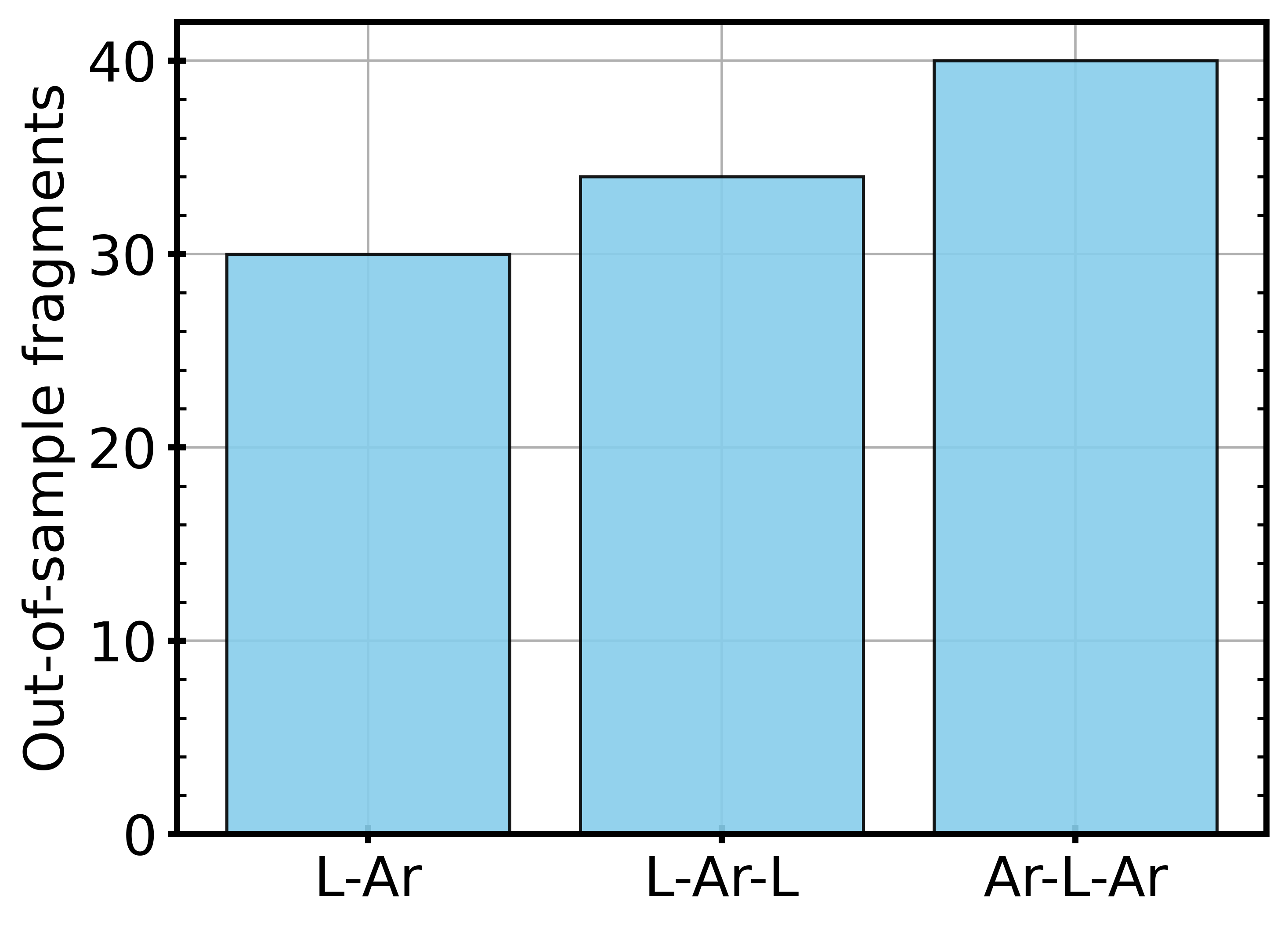}
    \caption{\textbf{Number of out-of-sample fragment occurrences during model validation.} An out-of-sample fragment occurrence is when the data set of polymers is split into training and test sets randomly, and there happens to be at least one polymer in the test set which is made up of at least one fragment that does not exist in the training set.}
    \label{SI_fig:OOTSFO}
\end{figure}

To accompany the results in the main text, we include additional information from the model validation stage. Fig.~\ref{SI_fig:OOTSFO} shows the number of out-of-sample fragment occurrences from the 50 different combinations of polymers in the training and test set during the repeated 5-fold CV. The various definitions of fragments result in different numbers of out-of-sample fragment occurrences: $L\-Ar$ has  fewer occurrences since there are fewer unique fragments, while $Ar\-L\-Ar$ has  the most occurrences.

\begin{figure}[H]
    \centering\includegraphics[width=.85\textwidth]{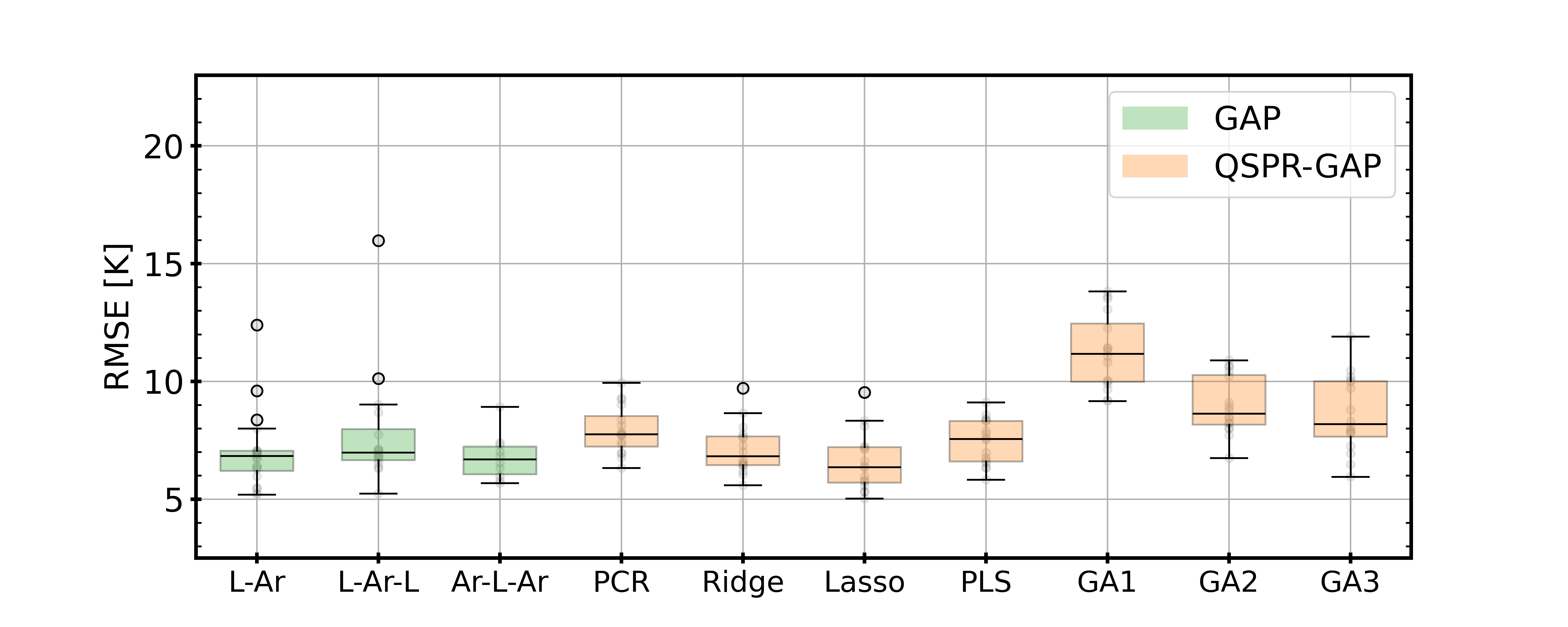}
    \caption{\textbf{Results of the QSPR-GAP model vs the GAP model with all out-of-sample fragment occurrences removed.} A repeated 5-fold cross validation scored by root-mean-squared error (RMSE). This corresponds to Fig.~2d in the main text with out-of-sample fragment occurrences removed. The orange and green boxes represent the interquartile range (IQR), covering the second and third quartiles (Q2 and Q3), with the black line indicating the median. Whiskers extend to the most extreme points within 1.5 times the IQR from Q1 and Q3, while outliers appear beyond this range.} 
\label{SI_fig:rmse_OOTSFO_removed}
\end{figure}

The comparative plots of QSPR-GAP vs GAP in the model performance evaluation (Fig.~2 of the main text) demonstrate how the GAP approach suffers for predictions of polymers containing out-of-sample fragment occurrences.  
Fig.~\ref{SI_fig:rmse_OOTSFO_removed} shows how the models compare when the training-test splits containing out-of-sample fragment occurrences have been removed. This is the same data presented in  Fig.~2d of the main text, except that the RMSEs calculated from training-test splits containing any out-of-sample occurrences have been removed from the plot.

\section{Fragment contributions $\hat{\beta}_i$}

The $\Tgval$ contributions $\hat{\beta}_i$ from each fragment $i$ is shown in Fig~\ref{SI_fig:beta_vals_all_fragments} for the following models: GAP, QSPR-GAP Lasso and QSPR-GAP GA2. The values of $\hat{\beta}_i$ are tabulated in Table~\ref{SI_table:beta_coefs_all_models} for all models applied to the $L\-Ar\-L$ fragments; the regressions were performed on the full data set of 146 polymers. The fragments are ordered from smallest to largest $\hat{\beta}_i$ values according to the QSPR-GAP GA2 model. Hyperparameters for the models PCR, Ridge, Lasso and PLS were determined from a 5-fold CV and
the hyperparameter which yielded the best performance (lowest RMSE) of the five folds was selected.
The optimal features (or descriptors) selected by the genetic algorithm was determined from the analysis in the main text. The hyperparameters and optimal features used to calculate the data in Fig.~\ref{SI_fig:beta_vals_all_fragments} are found in Table~\ref{SI_table:hyperparameters_final}.
\begin{table}[H]
    \centering
    \rowcolors{2}{white}{gray!25} 
    \begin{tabular}{>{\raggedright\arraybackslash}p{2cm}p{6cm}}
        \hline\hline
        \textbf{Estimator} & \textbf{Hyperparameters} \\
        \hline\hline
        PCR & $n_{\textrm{components}} = 13$ \\
        Ridge & $\alpha = 63.10$ \\
        Lasso & $\alpha = 0.007943$ \\
        PLS & $n_{\textrm{components}} = 7$ \\
        GA2 & descriptors = \texttt{Mor05m} and \texttt{Mor26m} \\
        GAP & N/A \\
        \hline\hline
    \end{tabular}
    \caption{\textbf{Hyperparameters for the models in Table~\ref{SI_table:beta_coefs_all_models}.}}
    \label{SI_table:hyperparameters_final}
\end{table}

\begin{figure}[H]
    \centering
    \begin{subfigure}{\textwidth}
        \centering
        \includegraphics[width=\textwidth]{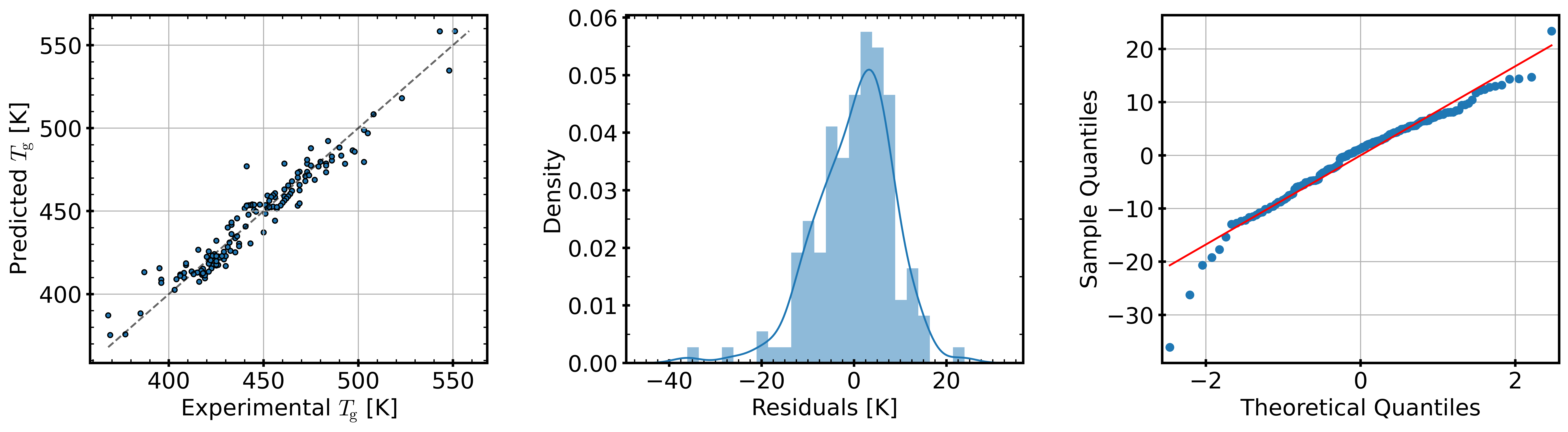}
        \caption{\textbf{QSPR-GAP GA2 model with Mor05m and Mor26m.}}
    \end{subfigure}
    \vspace{0.5cm}
    \begin{subfigure}{\textwidth}
        \centering
        \includegraphics[width=\textwidth]{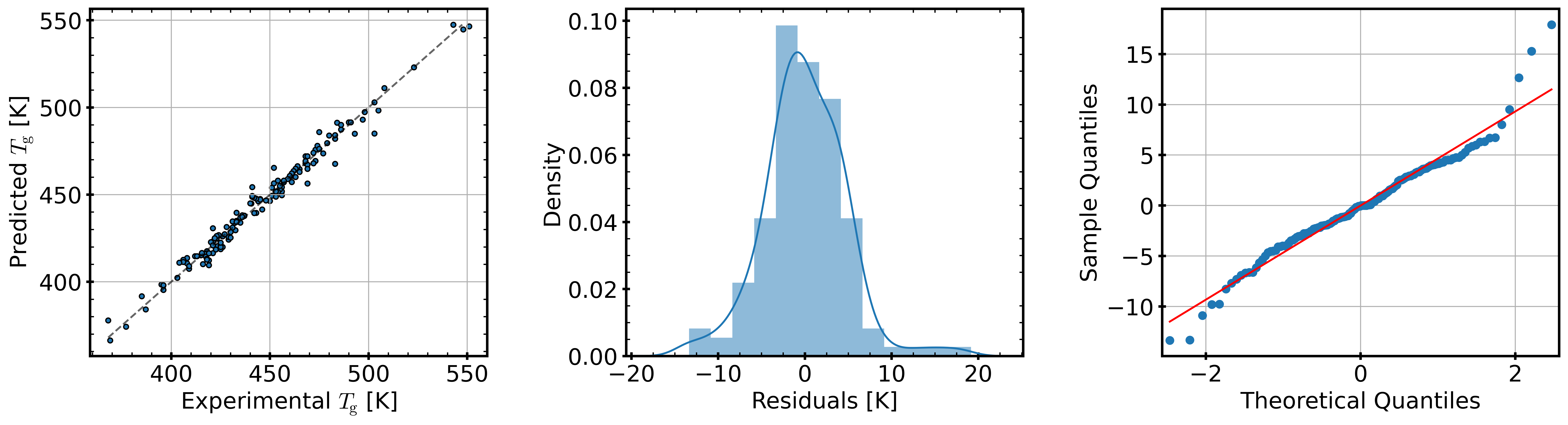}
        \caption{\textbf{GAP model.}}
    \end{subfigure}
    \caption{\textbf{Diagnostic plots for assumption of normality.} Each row of figures shows the fitted against measured $\Tgval$, the distribution of residuals, and Q-Q (quantile-quantile) plots of the residuals (from left to right) for: a)~QSPR-GAP GA2 model with \texttt{Mor05m} and \texttt{Mor26m}; the estimated standard deviation of the residuals is 8.5 K. b)~GAP model; the estimated standard deviation of residuals is 5.2 K.}
    \label{fig:normality_diagnostics}
\end{figure}

\begin{figure}[H]
    \centering\includegraphics[width=.45\textwidth]{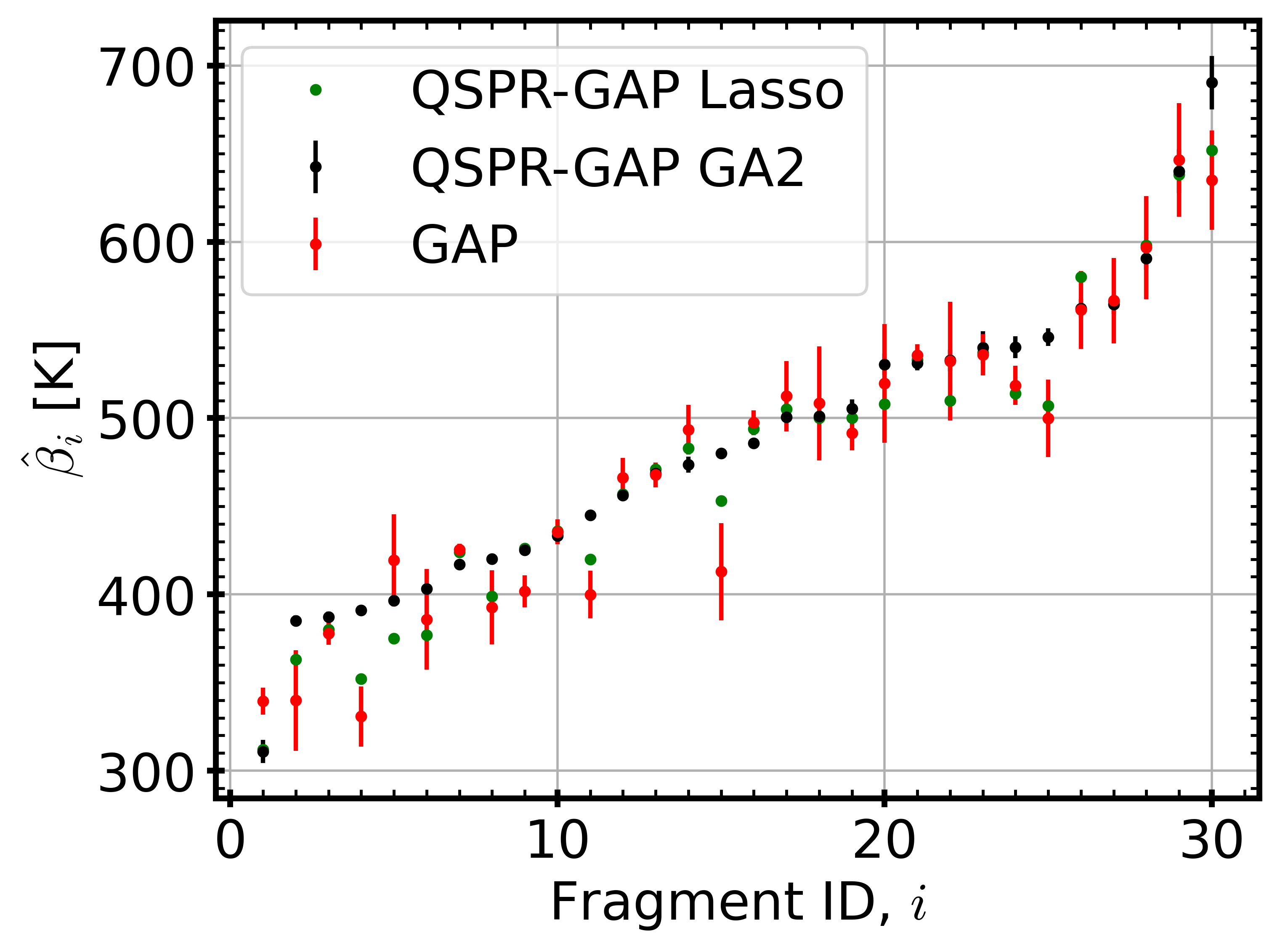}
    \caption{\textbf{Fragment $\Tgval$ contributions $\hat{\beta}_i$ for Lasso, GA2 ad GAP models.} $\hat{\beta}_i$ and corresponding confidence intervals are presented as indicated in Table~\ref{SI_table:beta_coefs_all_models}. Note here that the GAP model's confidence intervals (shown as vertical bars) are generally larger for the fragment identities that are poorly represented by the data in the fragment composition matrix; these are the the highly sparse columns in Fig.~\ref{SI_fig:composition_heatmap}, such as $i=15,i=20,$ and $i=28$.}
    \label{SI_fig:beta_vals_all_fragments}
\end{figure}

Diagnostic plots are shown in Fig.~\ref{fig:normality_diagnostics} to assess the assumption of normality for the residuals, which unless the sample size is sufficiently large, is required to report the inference in Table~\ref{SI_table:beta_coefs_all_models}, and in Table~I of the main text. The QSPR-GAP GA2 model with descriptors \texttt{Mor05m} and \texttt{Mor26m} shows a single extreme outlier with repeating unit structure: PENEPKDK \cite{Colquhoun2003FirstModeling} (see Fig.~\ref{SI_fig:all_groups_SI} for the corresponding chemical structure). The clear outlier in the plots has a Cook's distance of 0.12, and for further testing of it's influence, coefficient estimates and confidence intervals were determined from the data sample with the outlier removed.
Table~\ref{SI_table:regression_coefs_w_and_wo_outlier} shows the coefficient estimates and confidence intervals when the outlier is included in the fit (\textit{i.e.} fit to the full data sample of 146 polymers) and when the outlier is removed from the sample and fit to the remaining 145 polymers. Since the confidence intervals and parameter estimates are weakly influenced by this outlier, and given the large sample size (large number of observations per variable), the normality assumption of residuals is not required in this case. 
However, for the GAP model, the number of non-zero observations per predictor is not constant (see Fig.~\ref{SI_fig:composition_heatmap}); there are as few as one non-zero observation for certain predictors.
We report the confidence intervals for the GAP model in Table~\ref{SI_table:beta_coefs_all_models} however they should be interpreted with caution, given the slight deviation from linearity in the Q-Q plots shown in Fig.~\ref{fig:normality_diagnostics} at the extremes of the residuals.

\begin{table}[H]
    \centering
    \rowcolors{2}{gray!25}{white}
    \begin{tabular}{cccccc}   
        \hline\hline
        \rowcolor{white}
        \textrm{$\mu$} & \textrm{Descriptor} & \textrm{$\hat{\gamma}_\mu$ [\textrm{K}] } & \textrm{CI (95\%) L/U [K]} & \textrm{$\hat{\gamma}_\mu$ [\textrm{K}] } & \textrm{CI (95\%) L/U [K]} \\
        \rowcolor{white}
        & & incl. outlier & incl. outlier & w/o outlier & w/o outlier \\
        \hline
        \hline
        0 & \textit{--} & 298 & 286/310 & 296 & 284/307 \\
        1 & \texttt{Mor05m} & -58  & -67/-50 & -60  & -68/-52\\
        2 & \texttt{Mor26m} & -198  & -239/-157& -193  & -231/-155  \\
        \hline\hline
    \end{tabular}
    \caption{\textbf{Influence of the outlier on parameter estimations}. The table shows the QSPR-GAP GA2 model with descriptors \texttt{Mor01m} and \texttt{Mor26m} fit to the full data sample of 146 polymers; coefficient estimates and confidence intervals (CIs) are shown under `incl. outlier'. The table also presents this model fit to the data sample of 145 polymers with the outlier removed; coefficient estimates and CIs are shown under `w/o outlier'.}
    \label{SI_table:regression_coefs_w_and_wo_outlier}
\end{table}

\begin{table}[H]
    \centering
    \centering\includegraphics[width=.9\textwidth]{FigS3.pdf}
    \rowcolors{2}{gray!25}{white} 
    \resizebox{\textwidth}{!}{
    \begin{tabular}{>{\raggedright\arraybackslash}p{1.8cm}p{1cm}|p{1cm}|p{1cm}|p{1cm}|p{1cm}|p{1cm}p{1.5cm}p{1.5cm}p{1.5cm}|p{1.5cm}p{1.5cm}|p{2cm}}
        \hline\hline
        \multicolumn{2}{c|}{\textbf{Fragment ID}} & \textbf{PCR} & \textbf{Ridge} & \textbf{Lasso} & \textbf{PLS} & \multicolumn{4}{c|}{\textbf{GA2 Mor05m and Mor26m}} & \multicolumn{2}{c|}{\textbf{GAP}} & \textbf{Molar mass} \\
        Abbrv. & $i$ & $\hat{\beta}_i$ [K] & $\hat{\beta}_i$ [K] & $\hat{\beta}_i$ [K] & $\hat{\beta}_i$ [K] & $\hat{\beta}_i$ [K] & 95\% CI & Mor05m & Mor26m & $\hat{\beta}_i$ [K] & 95\% CI & $M_i$ [g/mol] \\\hline\hline
E-Tr-C & 1 & 319 & 326 & 312 & 321 & 311 & 304/317 & -0.511 & 0.0866 & 340 & 332/347 & 82 \\
E-m-E & 2 & 389 & 379 & 363 & 374 & 385 & 382/388 & -1.334 & -0.0459 & 340 & 311/368 & 92 \\
E-P-E & 3 & 378 & 385 & 380 & 377 & 387 & 384/390 & -1.374 & -0.0448 & 378 & 372/384 & 92 \\
E-m-C & 4 & 390 & 374 & 352 & 369 & 391 & 388/394 & -1.392 & -0.0589 & 331 & 314/348 & 91 \\
E-P-C & 5 & 371 & 376 & 375 & 367 & 396 & 394/399 & -1.506 & -0.0524 & 419 & 393/445 & 91 \\
E-o-E & 6 & 392 & 390 & 377 & 385 & 403 & 400/406 & -1.493 & -0.0899 & 386 & 357/414 & 92 \\
E-P-K & 7 & 420 & 420 & 424 & 422 & 417 & 415/419 & -1.687 & -0.1030 & 425 & 422/429 & 98 \\
E-m-K & 8 & 415 & 409 & 399 & 408 & 420 & 418/422 & -1.702 & -0.1149 & 393 & 372/414 & 98 \\
K-P-C & 9 & 411 & 412 & 426 & 414 & 425 & 424/427 & -1.762 & -0.1226 & 402 & 393/411 & 97 \\
E-P-d & 10 & 437 & 436 & 436 & 438 & 433 & 429/438 & -2.077 & -0.0708 & 436 & 428/443 & 112 \\
K-m-K & 11 & 436 & 434 & 420 & 427 & 445 & 443/447 & -1.899 & -0.1825 & 400 & 387/413 & 104 \\
K-P-K & 12 & 458 & 452 & 457 & 461 & 456 & 454/459 & -2.007 & -0.2071 & 466 & 455/477 & 104 \\
E-D-E & 13 & 478 & 474 & 471 & 477 & 469 & 465/472 & -2.374 & -0.1615 & 468 & 461/475 & 168 \\
E-D-C & 14 & 457 & 471 & 483 & 471 & 474 & 469/478 & -2.482 & -0.1551 & 493 & 479/508 & 167 \\
E-N-E & 15 & 448 & 464 & 453 & 450 & 480 & 478/482 & -2.292 & -0.2438 & 413 & 385/440 & 142 \\
E-P-S & 16 & 496 & 494 & 494 & 498 & 486 & 484/488 & -2.381 & -0.2462 & 497 & 490/504 & 116 \\
E-qN-E & 17 & 491 & 496 & 505 & 483 & 500 & 498/503 & -2.488 & -0.2887 & 512 & 493/532 & 142 \\
E-rN-E & 18 & 495 & 497 & 500 & 485 & 501 & 498/504 & -2.472 & -0.2965 & 508 & 476/541 & 142 \\
E-D-K & 19 & 502 & 499 & 500 & 502 & 505 & 500/510 & -2.808 & -0.2188 & 492 & 482/501 & 174 \\
K-Dm-K & 20 & 517 & 531 & 508 & 516 & 531 & 524/537 & -3.084 & -0.2647 & 520 & 486/553 & 180 \\
E-rN-K & 21 & 533 & 533 & 533 & 532 & 531 & 527/535 & -2.852 & -0.3359 & 536 & 530/542 & 148 \\
K-mDp-K & 22 & 509 & 525 & 510 & 514 & 533 & 527/539 & -3.096 & -0.2724 & 532 & 499/566 & 180 \\
K-N-K & 23 & 545 & 544 & 537 & 550 & 540 & 531/549 & -2.543 & -0.4726 & 536 & 524/548 & 154 \\
K-D-K & 24 & 513 & 515 & 514 & 509 & 540 & 534/546 & -3.166 & -0.2891 & 519 & 508/530 & 180 \\
K-Do-K & 25 & 509 & 518 & 507 & 489 & 546 & 541/551 & -3.118 & -0.3326 & 500 & 478/522 & 180 \\
K-rN-K & 26 & 579 & 574 & 580 & 590 & 562 & 557/568 & -3.150 & -0.4058 & 561 & 539/583 & 154 \\
K-qN-K & 27 & 560 & 567 & 566 & 574 & 564 & 559/570 & -3.242 & -0.3900 & 567 & 542/591 & 154 \\
E-rN-S & 28 & 608 & 592 & 598 & 593 & 591 & 584/597 & -3.441 & -0.4637 & 597 & 568/626 & 166 \\
S-D-S & 29 & 630 & 640 & 638 & 623 & 640 & 628/652 & -4.398 & -0.4306 & 646 & 614/679 & 216 \\
S-N-S & 30 & 654 & 657 & 652 & 651 & 690 & 675/705 & -3.925 & -0.8238 & 635 & 607/663 & 190 \\
        \hline\hline
    \end{tabular}}
    \caption{\textbf{Table of $\hat{\beta}_i$ for all fragments and models.}
    The first column provides the fragment identity as shown in the figure above. Here, we tabulate the resulting $\hat{\beta}_i$ values
    from the following QSPR-GAP methods: Principal component regression (PCR), Ridge regression, Lasso regression, Partial Least Squares (PLS) regression, Ordinary Least Squares regression with a 2 feature subset selection via the genetic algorithm (GA2) for the two 3D-MoRSE descriptors noted. A GAP model of the form $L\textrm{-}Ar\textrm{-}L$ is also presented, using OLS. Upper (UCI) and lower (LCI) 95\% confidence intervals for the unbiased estimators (GA2 and GAP) are also presented. Note that the molar mass $M_i$ of an $L\textrm{-}Ar\textrm{-}L$ fragment is the molar mass of half of each $L$ group and the full $Ar$ group: $M_i=M_{L_{i1}}/2 + M_{Ar_i} + M_{L_{i2}}/2$.
    }
\label{SI_table:beta_coefs_all_models}
\end{table}

\section{Atom pair contributions $\hat{\pi}_{ij}$}

\begin{figure}[H]
    \centering    
    {\includegraphics[width=0.85\textwidth]{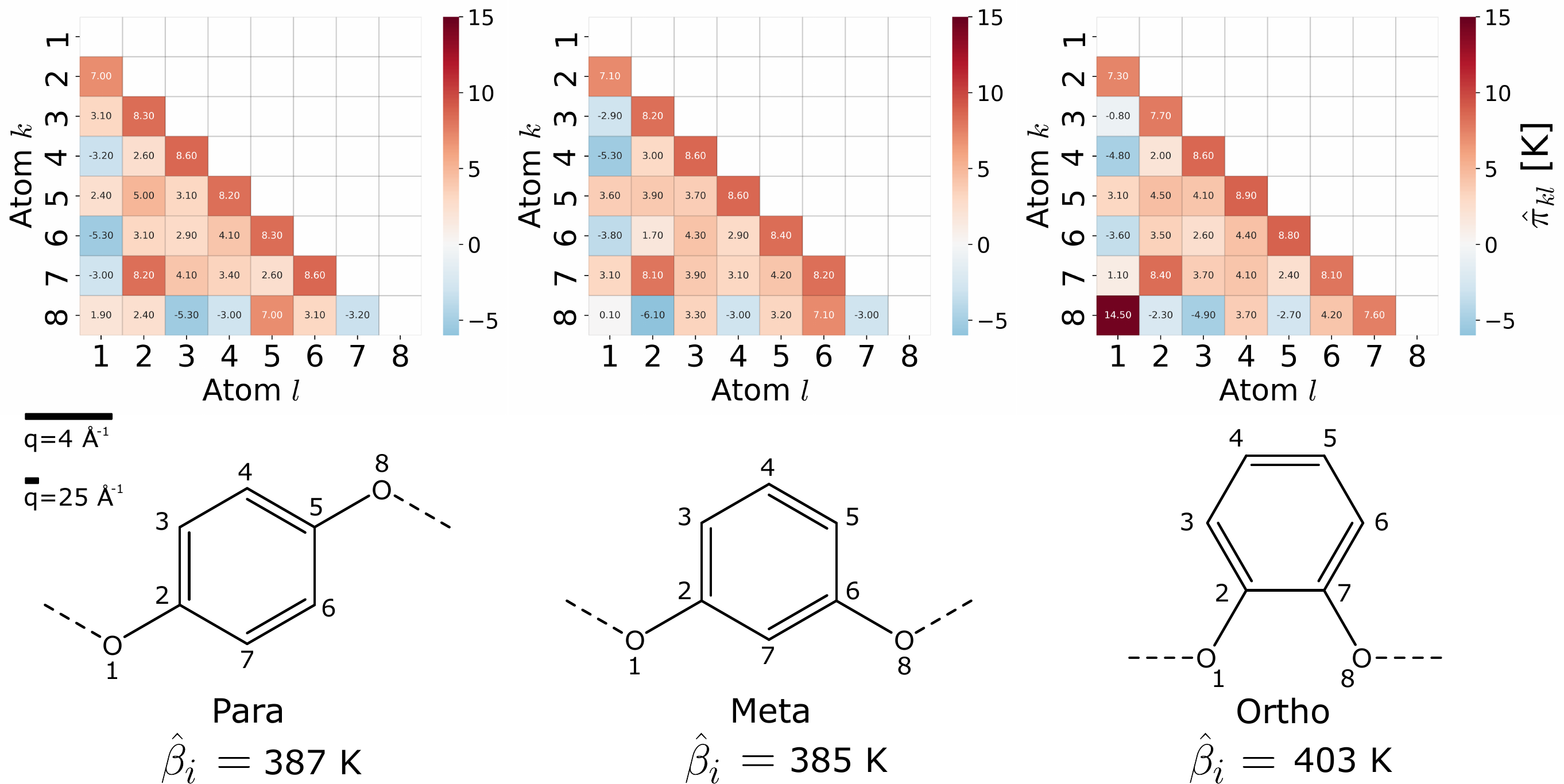}}
    \caption{\textbf{Atomic pair contributions for phenylene variants.}}
    \label{SI_fig:phenyls_pair_contributions}
\end{figure}

\begin{sidewaysfigure}
    \centering
    \includegraphics[width=1\textwidth]{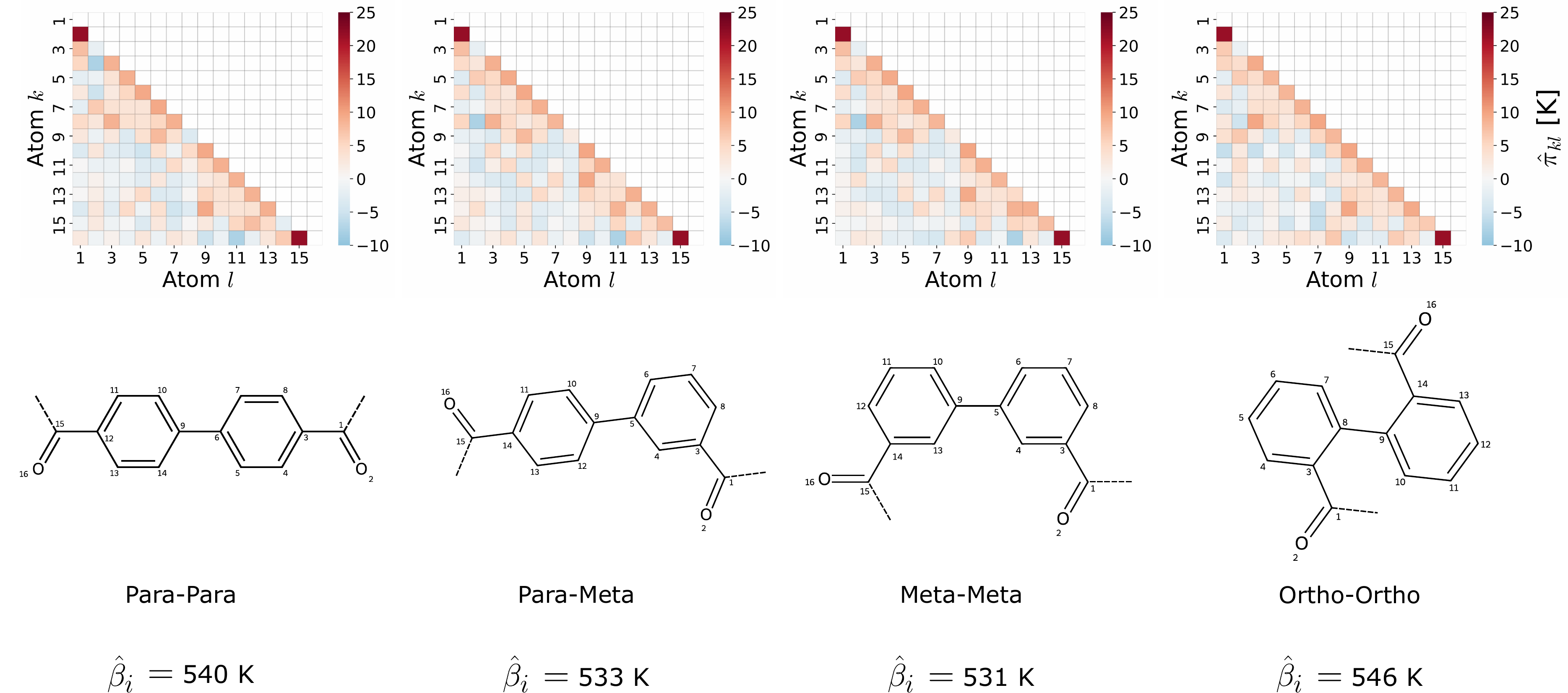}
    \caption{\textbf{Atomic pair contributions for biphenylene variants.}}
    \label{SI_fig:diphenyls_pair_contributions}
\end{sidewaysfigure}

\begin{figure}[H]
\begin{center}
\includegraphics[width=\textwidth]{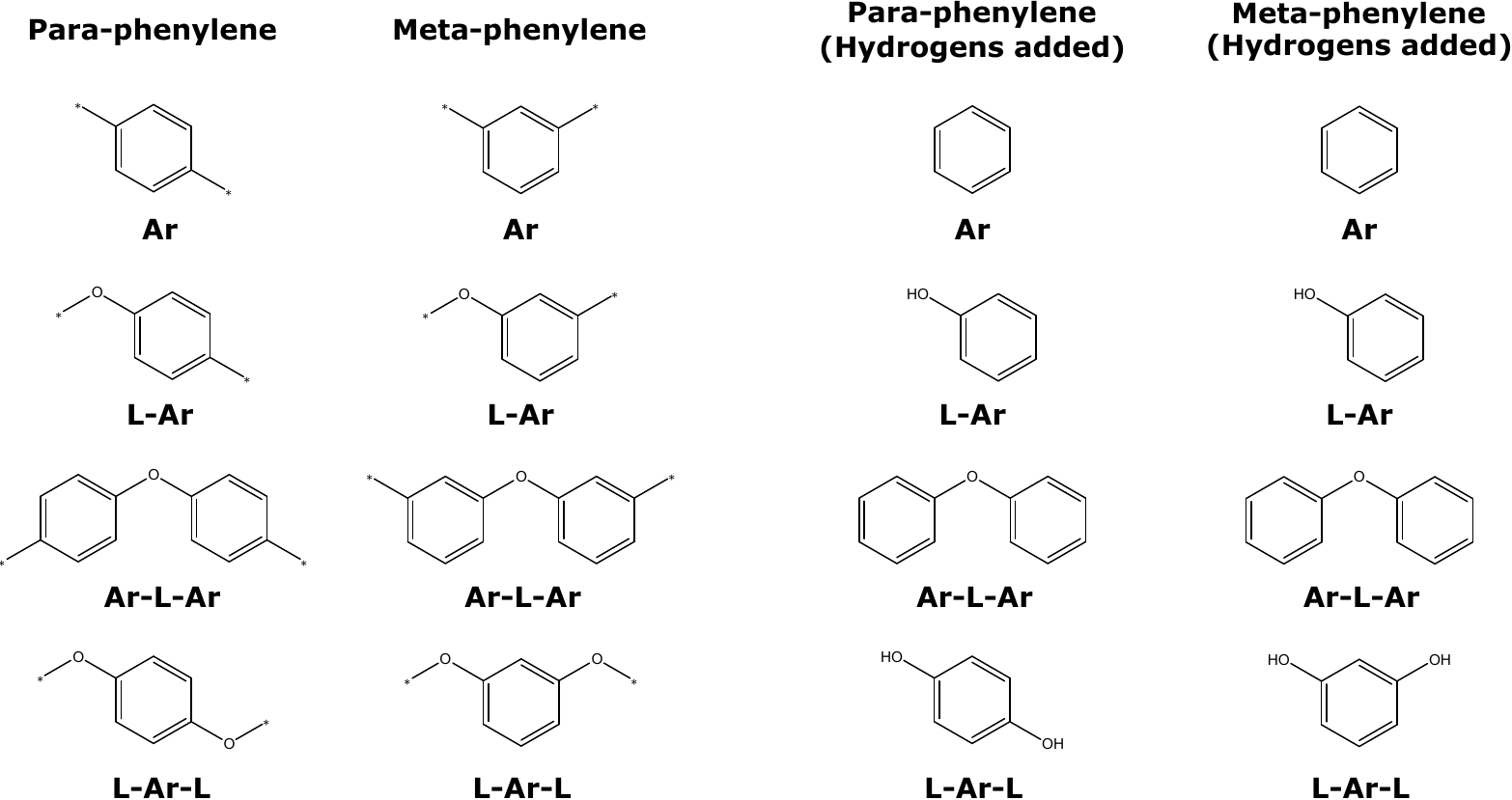}
\end{center}
\caption{\textbf{Motivating the choice of} $\boldsymbol{L\textrm{-}Ar\textrm{-}L}$\textbf{.} The figure shows two examples of phenylene variants (meta and para linked) that appear in different fragments. The connection of each fragment to its neighbours (in the backbone) is through the symbol $-\ast$. We  must replace $-\ast$ in each fragment by hydrogens in order to generate their 3D descriptors. When performing this replacement for para and meta phenylenes, the only definition that uniquely distinguishes these two is the $L\textrm{-}Ar\textrm{-}L$ definition. Hence, we use the $L\textrm{-}Ar\textrm{-}L$ definition of a fragment.}
\label{SI_fig:LRL_motivation}
\end{figure}

The following results in Fig.~\ref{SI_fig:phenyls_pair_contributions} and Fig.~\ref{SI_fig:diphenyls_pair_contributions} are the estimated atomic pair $\Tgval$ contributions, denoted $\hat{\pi}_{kl}$ for each $k$th and $l$th atom pair. The fragments shown are a selected few from the 30 unique fragments in the dataset.
The contributions were calculated from the two descriptors \texttt{Mor05m} and \texttt{Mor26m} selected by the GA (Eq.~7 in the main text). The overall fragment contribution, denoted $\hat{\beta}_i$ is determined by summing over all the $\hat{\pi}_{kl}$ values that exist in fragment $i$ plus the constant $\hat{\gamma}_0$ (Eq.~6 in the main text). In these plots, $\hat{\pi}_{kl}$ contributions from atom pairs containing a hydrogen are ignored since these are very weak contributions.


\newpage
\section{Fragment definitions}

\newpage



%